\documentclass[journal]{IEEEtran}
\usepackage{amsmath,amsfonts}
\usepackage{algorithmic}
\usepackage{algorithm}
\usepackage{array}
\usepackage{amsthm}
\usepackage{amssymb}
\usepackage{textcomp}
\usepackage{stfloats}
\usepackage{url}
\usepackage{verbatim}
\usepackage{graphicx}
\usepackage{cite}

\usepackage{booktabs}
\usepackage{multirow}
\usepackage{makecell}
\usepackage[caption=false, font=footnotesize]{subfig}
\usepackage{xcolor}

\newtheorem{theorem}{Theorem}

\newtheorem{assumption}{Assumption}

\begin{document}
	\title{Decision Uncertainty-Oriented Watermarking (DUOW): Securing Multi-Agent Routing Against Policy Cloning}
    
	\author{ Ziming Zhang, Li Li, Guorui Feng, Hanzhou Wu, Xinpeng Zhang, and Athanasios V. Vasilakos

        \thanks{Li Li (email: llichn@shu.edu.cn) and Xinpeng Zhang (email: xzhang@shu.edu.cn) are the corresponding authors.}
        \thanks{Ziming Zhang, Li Li, Guorui Feng, Hanzhou Wu, and Xinpeng Zhang are with the School of Communication and Information Engineering, Shanghai University, Shanghai 200444, China (email: zhangzmoli@shu.edu.cn).}
        \thanks{Athanasios V. Vasilakos is with the Department of ICT and Center for AI Research, University of Agder (UiA), Jon Lilletuns vei 9, 4879 Grimstad, Norway (email: th.vasilakos@gmail.com).}
    
        }

    \markboth{IEEE TRANSACTIONS ON INFORMATION FORENSICS AND SECURITY, VOL. XX, NO. XX, 2026}%
    {Shell \MakeLowercase{\textit{et al.}}: A Sample Article Using IEEEtran.cls for IEEE Journals}

	\maketitle

\begin{abstract}
Routing policies in Multi-Agent Systems (MAS) represent valuable intellectual property, yet their exposure through execution logs creates significant risks of unauthorized policy cloning. Existing agent watermarking methods protect intellectual property by modifying internal generation processes, such as perturbing token sampling distributions or injecting structural mutations into action trajectories, which may compromise task fidelity or suffer from limited transferability across heterogeneous platforms. To address these limitations, we propose Decision Uncertainty-Oriented Watermarking (DUOW), a lightweight framework that operates at the external decision layer without modifying model parameters or generated outputs. DUOW introduces a plug-and-play scoring mechanism that evaluates candidate agents from semantic and operational perspectives, identifies functionally redundant routing decisions, and selectively redirects choices among equivalent alternatives to embed ownership signals. By coupling watermark behaviors with utility-preserving routing decisions, DUOW enables cloned policies to retain ownership signals during policy imitation while increasing the difficulty of watermark removal. A confidence-driven adaptive threshold dynamically determines embedding opportunities, balancing watermark capacity and task fidelity under diverse routing contexts. Extensive evaluations on four multi-agent routing benchmarks demonstrate that DUOW preserves high task utility, achieves reliable ownership verification under up to 90\% log erasure, and introduces negligible computational overhead, with an average per-step latency of 0.043 ms and total embedding overhead below 0.54 ms.
\end{abstract}


\section{Introduction}
\IEEEPARstart{R}{ecent} advances in Large Language Models (LLMs) have enabled the development of autonomous agents that can perceive, plan, and execute tasks with minimal human intervention. However, single-agent architectures often struggle with complex industrial tasks due to limited scalability and resource capacity \cite{11218933,hendrycks2020measuring,patil2024gorilla}. To address this issue, most agent systems are increasingly adopting Multi-Agent Systems (MAS) \cite{hong2024metagpt,acharya2025agentic,wang2024survey,xi2025rise}. MAS improves efficiency by coordinating heterogeneous agents, specialized tools, and external API resources. In such systems, routing policies play a critical role because they determine which agent should execute a task under dynamic conditions \cite{jiang2023llm-blender}. Designing effective routing policies requires extensive algorithmic optimization and large-scale computational resources. Consequently, routing policies have become a core form of intellectual property (IP) in modern AI systems.   

\begin{figure}[!t]
    \centering
    \includegraphics[width=\linewidth]{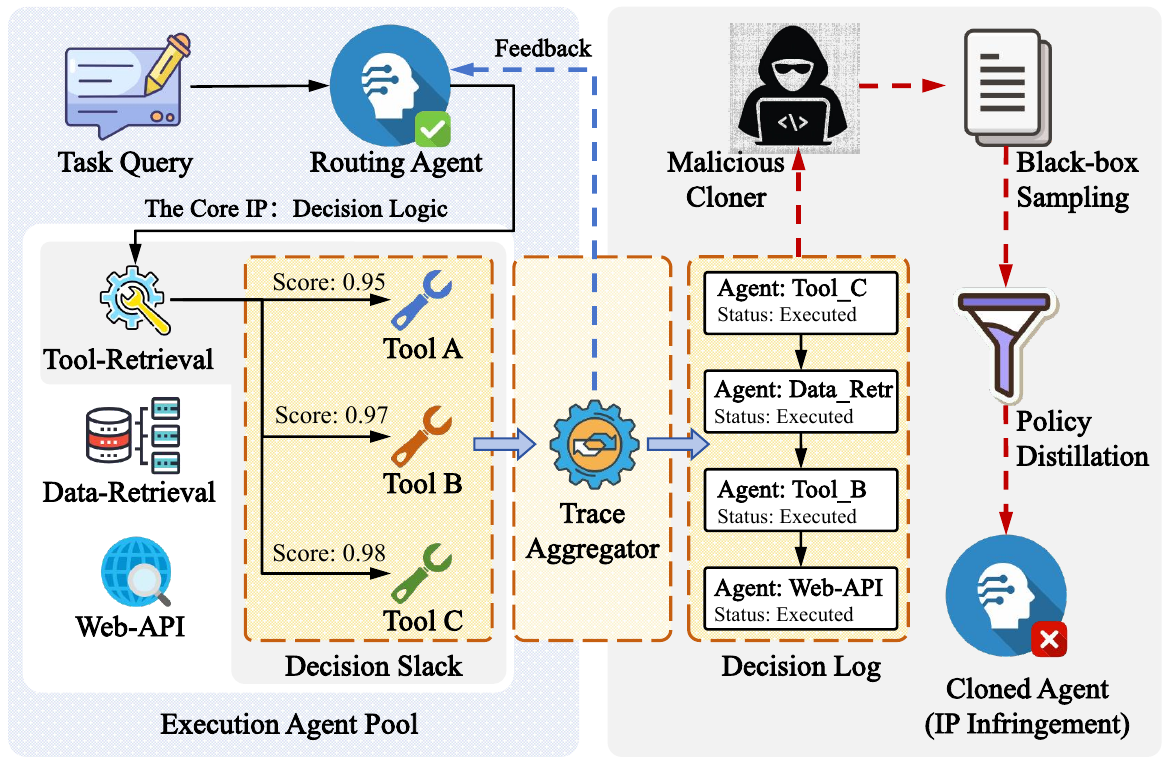}
    \caption{\textbf{Threat Model of Policy Cloning in MAS.} The routing agent selects the most suitable execution subagent from the execution agent pool. The routing result is then recorded in the decision log containing the execution agent IDs and execution statuses. Attackers may acquire unauthorized routing agent replicas via insider data breaches, or knowledge distillation from public APIs. The unauthorized replica violates legitimate rights and interests.}
    \label{fig:threat}
\end{figure}

However, in practical deployments\cite{deng2025ai}, proprietary routing policies in MAS are exposed to potential leakage and malicious attacks \cite{10976441,zhao2026shadowcot}. MAS often need to record and externalize execution logs and scheduling traces for purposes such as billing, performance monitoring, and system debugging. While these observable behavior records enhance system transparency, they also inadvertently expose the system’s core decision-making processes. Attackers can analyze these execution sequences to infer routing logic and perform black-box policy extraction or malicious imitation attacks \cite{liu2025evaluating,li2025differentiation,zhao2025survey,gudibande2024false}.
Existing studies have shown that attackers can achieve policy cloning at minimal cost solely by observing agent behaviors \cite{tramer2016stealing}. Gudibande et al. \cite{gudibande2024false} further demonstrate that, although black-box imitation learning struggles to replicate the deep reasoning capabilities of a model, it is highly effective at reproducing endogenous behavioral tendencies and execution trajectories. 
As a result, adversaries can closely replicate proprietary scheduling patterns and deploy infringing systems with similar functionality. These vulnerabilities threaten the commercial value of MAS and underscore the critical need for robust intellectual property protection mechanisms for proprietary routing policies.
\subsection{Motivation: Watermarking Beyond Token and Action}
 As a widely adopted approach for intellectual property protection \cite{10843740, tang2026cref, gao2026pathmark}, model watermarking embeds verifiable ownership identifiers into model outputs to support copyright verification and provenance tracing.
For LLMs, a range of strategies have been explored—from biasing token sampling distributions \cite{kirchenbauer2023watermark,kuditipudi2023robust} to embedding semantics-invariant signatures \cite{liu2024semantic} and multi-bit payloads \cite{chang2024postmark,guan2024codeip}. Despite their diversity, such text-based watermarking paradigms become ineffective in autonomous agent environments. Unlike natural language, which thrives on extensive semantic redundancy, autonomous agents translate high-level policies into discrete API actions under rigid control-flow dependencies and restricted vocabulary boundaries. Consequently, imposing token-level probability perturbations inevitably disrupts the deterministic execution logic, causing cascading parsing failures.
Moreover, MAS's closed-loop planning propagates token perturbations through feedback, whose stepwise bias amplification causes long-range drifts that undermine end-to-end task fidelity.

To circumvent the inherent threat to task fidelity posed by token-level manipulation, recent studies have extended watermarking from token perturbations to the behavioral dimension of autonomous agents. These approaches roughly fall into two categories: probability-centric methods and action-sequence-based methods. Probability-centric methods embedding watermarks by modulating behavioral distributions. Representative methods such as Agent Guide \cite{huang2025agent} and AgentMark \cite{huang2026agentmark} impose explicit biases or distribution-preserving sampling constraints. These methods rely on explicit access to internal candidate distributions during verification, which becomes infeasible under black-box API deployments. Action-sequence-based methods exploit structural or functional redundancies within execution paths, which inject structural mutations into tool execution paths \cite{wang2026protecting} or embed hook actions into sequential trajectories\cite{meng2026watermarking}. These methods support black-box verification, but their path-equivalence mechanisms are tied to static tool definitions, hampering transferability across heterogeneous settings or dynamic endpoint changes.

As both families of existing methods share a fundamental flaw: they treat watermarking as an exogenous perturbation imposed upon the agent’s output distribution or action sequence, rather than as an inherent property of the decision‑making logic. This forces them either to depend on internal state access (which is unavailable in black‑box deployments) or to rely on brittle, platform‑specific execution paths that break under endpoint changes or heterogeneous environments. Moreover, because the watermark is added as a superficial layer, it can be easily washed out during policy distillation or imitation learning, where the student model sees only the final behavioral traces and abstracts away the perturbation patterns. These deficiencies motivate a fundamentally different perspective—one that exploits the structural redundancy already present in the routing process itself. In multi‑agent systems, multiple candidate agents often possess overlapping capabilities, yielding multiple near‑equivalent routing decisions for a given subtask. Rather than treating watermarking as an external modification, we propose to embed ownership signals directly into the selection among these functionally redundant alternatives. This decision‑level paradigm decouples the watermark from any internal model parameters or generated content, enabling robust black‑box verification while preserving task utility.

\subsection{Main Contributions}
We introduce Decision Uncertainty‑Oriented Watermarking (DUOW), a lightweight framework that operates at the external decision layer. DUOW evaluates candidate agents through a plug‑and‑play scoring mechanism that jointly considers semantic alignment and operational utility, identifies routing steps with sufficient decision slack, and selectively redirects the choice among equivalent candidates to encode ownership bits. By coupling the watermark with utility‑preserving routing decisions, DUOW ensures that cloned policies inevitably inherit the watermark during imitation, making removal substantially more difficult—all without modifying model internals or compromising end‑to‑end task performance.

\begin{itemize}
    \item We propose a decision-level watermarking framework Decision Uncertainty-Oriented Watermarking.
    Different from existing token or action level, DUOW embeds ownership signals into redundant routing decisions rather than generated content or action, enabling black-box verification from decision traces.
    \item We design a plug‑and‑play scoring method to facilitate the identification of watermark-compatible routing paths. Rather than selecting only the optimal sub-agent as in traditional scoring paradigms, the proposed method assigns scores to all candidate sub-agents by jointly considering static semantic alignment and dynamic operational utility, thereby identifying substitutable routing paths that can accommodate watermark embedding. 
    \item We design an adaptive threshold mechanism based on the proposed agent scoring scheme to dynamically estimate decision slack. By jointly considering candidate competitiveness and routing preference, the proposed mechanism identifies substitutable routing opportunities for watermark embedding while minimizing the impact on task performance, overcoming the limitations of fixed thresholds that fail to adapt to varying levels of decision slack.
    \item We validate the effectiveness of DUOW through extensive evaluations across four heterogeneous multi-agent routing domains, showing its robustness in watermark inheritance against policy distillation and cloning, reliable recovery under log truncation, and negligible impact on task utility.
\end{itemize}

\section{Related Work}
\subsection{Multi-Agent Systems and Routing Policies}
Multi-Agent Systems (MAS) coordinate multiple specialized agents to solve complex problems through task decomposition and dynamic allocation. The routing agent serves as the central coordinator of MAS, responsible for dynamically assigning subtasks to the most suitable execution endpoint based on the current context. Early efforts on the agent side focused on improving the capabilities of LLM-based agents in interacting with external tools and environments. Yao et al.~\cite{yao2022react} introduced the ReAct paradigm, which interleaves reasoning traces with action execution, allowing agents to plan, observe, and adjust their behavior based on environmental feedback. Schick et al.~\cite{schick2023toolformer} further extended this capability by enabling LLMs to autonomously invoke APIs through token-level execution commands. Building on these advances, Qin et al.~\cite{qin2024toolllm} formulated ToolLLM, scaling agent capabilities to orchestrate thousands of real-world APIs across complex multi-step and cross-tool workflows. These works collectively expanded the scope and complexity of tasks that MAS can handle through agent collaboration.

As agent capabilities matured, the number of available tools and APIs expanded rapidly. The critical challenge thus shifted from enabling tool use to deciding which tool to invoke at each step, raising the question of how to make the right routing decision under varying task contexts. Chen et al.~\cite{chen2023frugalgpt} introduced FrugalGPT, which routes queries through an adaptive cascade of LLMs based on task difficulty, improving decision accuracy while reducing cost. Zhai et al.~\cite{zhai2025enhancing} refined this to step-level granularity by training a lightweight Q-value model for adaptive action selection, enabling more precise routing in multi-step workflows. More recently, the OPTIMA framework \cite{chen2025optima} employs Monte Carlo Tree Search and direct preference optimization to learn routing policies that minimize black-box communication costs without relying on heuristic rules. These works collectively demonstrate that optimizing routing decisions is central to maximizing overall system performance in MAS.

\subsection{Agent Watermarking}
Agent systems face various intellectual property threats, including the replication of behavioral patterns, strategic reasoning processes, and decision-making logic without authorization. Among these intellectual property threats, unauthorized replication of routing policies is of particular concern, as routing policies govern task allocation and largely determine overall system performance. Protecting routing agents against policy theft and unauthorized cloning has therefore attracted growing interest, prompting a range of watermarking techniques that embed ownership signals at different levels of agent behavior.
Some approaches operate at the probability level. Agent Guide \cite{huang2025agent} and AgentMark \cite{huang2026agentmark} embed watermarks by modulating time-varying behavioral distributions. AgentMark attempts black-box deployment via distribution elicitation, but still requires explicit access to logit profiles or introduces high computational overhead through multi-round sampling. Moreover, because these identifiers are encoded as subtle shifts in probability lists, they are easily smoothed out by student models as stochastic noise during policy distillation. Other approaches operate at the action level. AgentWM \cite{wang2026protecting} exploits semantic equivalence among alternative tool execution paths and biases the agent towards specific realizations to encode watermark information. ActHook \cite{meng2026watermarking}, in contrast, inserts auxiliary hook actions into sequential trajectories while preserving the original task outcome. 
 
Despite their diversity, these existing methods share a common root limitation: they all embed ownership as external perturbations: probability-level biases wash out under distillation, while action-level mechanisms are tied to static definitions and lack transferability. More fundamentally, none exploits the functional redundancy naturally present among candidate agents in multi-agent systems, where multiple near-equivalent routing choices exist for a given subtask. This untapped redundancy points toward a fundamentally different watermarking paradigm: embedding ownership into the selection among equivalent alternatives rather than perturbing generated content or execution traces, potentially offering robustness against cloning and log truncation while preserving task fidelity.

\section{Threat Model}
To set the stage for our research motivation, we first outline the basic MAS architecture and then derive the threat model for our proposed method accordingly.
\subsection{Multi-Agent System (MAS)}
In MAS, the routing agent receives a user query, decomposes it into sequential subtasks, and dispatches each subtask to an appropriate execution endpoint selected from a pool of candidate agents, tools, or APIs. To support execution auditing and system debugging, the entire routing process is recorded as chronological decision logs. As illustrated in Fig.~\ref{fig:threat}, MAS involves three core architectural entities:
  
\subsubsection{Execution Agent Pool ($\mathcal{A}$)} Represents a massive library that contains a wide variety of downstream sub-agents, third-party APIs, and specialized programmatic tools. For each subtask, the routing agent selects an appropriate execution target from $\mathcal{A}$ according to the task requirements and the capabilities of candidate agents. Under different task contexts, multiple candidate agents often possess overlapping functions or competitive proficiencies, providing alternative paths to accomplish the same subtask.

\subsubsection{Routing Agent ($R$)} Serves as the central decision-making entity that selects the most suitable execution endpoint for each subtask based on the current context and historical trajectory. Given a user request $Q$, the agent decomposes it into a sequence of subtasks. At each step $t$, it generates a localized query $q_t$ from the accumulated history and feeds it to the underlying LLM, which produces a logit vector over the candidate pool $\mathcal{A}$:
\begin{equation}
    \mathbf{z}_t = [z_{1,t}, \dots, z_{n,t}] \in \mathbb{R}^n,
\end{equation}
where each logit reflects the model's preference for the corresponding agent. The routing LLM is typically trained with a cross-entropy loss that maximizes the probability of the ground-truth candidate. In native operation, the router adopts a greedy policy:
\begin{equation}
    a_t = \pi_0(q_t) = \arg\max_{a_i \in \mathcal{A}} z_{i,t}.
\end{equation}
The decision $(q_t, a_t)$ and the resulting observation are recorded in the trajectory history for subsequent steps.

\subsubsection{Decision Log ($\mathcal{L}$)} {In practical deployments, decision logs are made accessible to users, platform operators, or third-party services for auditing, billing, and debugging purposes. It} serves as a historical record that stores the execution traces of the system. Specifically, it aggregates the planning trajectories by chronologically tracking each input query, request timestamp, and the corresponding execution agent ID selected by the routing agent.

\begin{figure*}[!t]
    \centering
    \includegraphics[width=\linewidth]{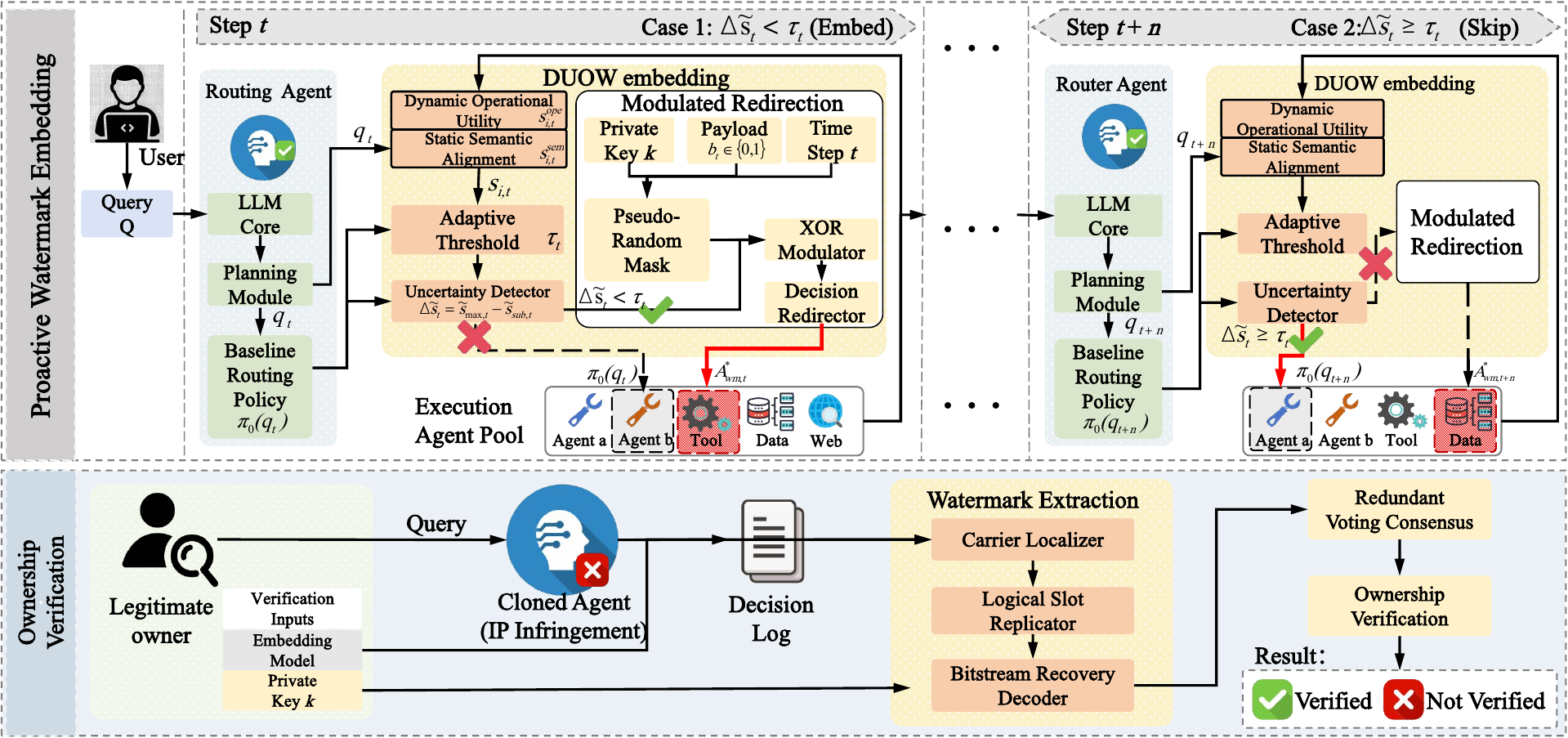}
    \caption{\textbf{Framework of DUOW.} \textbf{Watermark Embedding:} DUOW embeds the watermark into routing decisions among functionally equivalent agents during sequential decision-making in MAS, progressively accumulating a multi-bit payload through distributed insertion over multiple routing steps. \textbf{Ownership Verification:} The legitimate owner extracts the bitstream from observable decision traces and recovers the full payload through redundant voting consensus.}

    \label{fig:pipeline}
\end{figure*}

\subsection{Adversary}
The adversary seeks to obtain the target routing agent or a functionally equivalent replica without incurring the substantial cost of developing and optimizing the routing policy from scratch.
The adversary may directly acquire the target routing agent or its model parameters through model leakage, insider disclosure, or other forms of unauthorized access. When direct access is unavailable, the adversary can collect observable task–routing behaviors through public service interfaces, leaked execution logs, monitoring records, interaction histories, or other accessible system outputs, and subsequently trains a surrogate routing model using knowledge distillation or other imitation learning techniques.

\subsection{Defense Model}
To prevent the aforementioned attacks, the routing agent owner can embed copyright identifiers during model development. The knowledge and capabilities of the owner are considered in two stages:

\subsubsection{Watermark Embedding} 
As the legitimate developer of the original routing agent, the owner has full control over the model prior to deployment and can manipulate its routing decisions to embed ownership information. At the end of this phase, the owner retains the secret watermarking key and the embedding configurations required for subsequent ownership verification.

\subsubsection{Copyright Verification} Due to commercial trade secrets or legal compliance boundaries, the owner may be unable to access internal information of the suspect model during copyright verification, including parameters, architecture, and intermediate layer outputs. Therefore, we assume that ownership verification is conducted under a black‑box setting, where the owner can only send queries to the suspect system's public API and audit the returned sequential decision logs.

\section{Methods}
\label{sec:Methods}
In a multi‑agent system, the routing agent dynamically selects the most suitable sub‑agent based on the current user query, historical decision trajectory, subtask decomposition results, and the capability descriptions. Since multiple sub‑agents often exhibit nearly equivalent performance on the same task, the routing agent can choose among several competing candidates, giving rise to decision redundancy that naturally facilitates watermark embedding. The core idea of Decision Uncertainty‑Oriented Watermarking (DUOW) is to exploit this margin by encoding ownership information into the selection of alternative paths, thereby embedding the watermark into the multi‑agent routing process.

Current routing agents typically select only the optimal sub‑agent while disregarding other candidates, thus failing to identify functionally equivalent alternatives and lacking the redundancy necessary for watermark embedding. The routing agent LLM is typically trained with a cross-entropy loss, as shown in Eq. \eqref{eq-loss}:
\begin{equation}
   L = -\log(p_{y,t}) ,
   \label{eq-loss}
\end{equation}
where $p_{y,t}$ denotes the predicted probability of selecting the optimal target $y$ at routing step $t$. It could be seen that the cross-entropy loss only maximizes the probability of the top‑1 choice while neglecting others. As a result, the logits associated with the remaining options do not reflect their corresponding downstream agents' functional competitiveness. Therefore, directly exploiting the original routing policy fails to yield usable decision slack. To address this issue, DUOW introduces a score module that employs a reference scoring mechanism to compute reliable scores for all candidate agents. This module is plug‑and‑play and readily adaptable to any multi‑agent system.

Based on this score module, we can define the decision slack—the redundancy space among functionally equivalent agents. Considering not every routing step in MAS offers similarly capable downstream agents, resulting in certain steps provide insufficient slack for embedding. Therefore, we design an adaptive embedding mechanism that assesses feasibility step by step. As shown in Fig.~\ref{fig:pipeline}, embedding is carried out stepwise along the routing trajectory. If a step lacks sufficient uncertainty or viable alternatives, it is skipped, and the 1‑bit insertion is deferred to the next step. The details are as follows.

\subsection{Plug-and-Play Score Module}
We construct a scoring module that evaluates each candidate agent from two complementary perspectives: static semantic alignment and dynamic operational utility. The semantic alignment score measures the degree to which an agent's documented capability matches the current task requirement, while the operational utility score captures runtime conditions such as response latency and current workload.
These two dimensions jointly provide a reference score for estimating the suitability of each candidate agent for the current routing step.

\subsubsection{Static Semantic Alignment}
To measure the functional alignment between a candidate agent and the current subtask, we transform both the user query $q_t$ and the agent's capability descriptor $e_i$ into the same feature space using a pre-trained semantic encoder.
Then we quantify their similarity by the static semantic alignment score $s_{i,t}^{\text{sem}}$, which is computed as the scaled cosine similarity:
\begin{equation}
    \label{eq:semantic_alignment}
    s_{i,t}^{\text{sem}} = \sigma \left( \frac{\Phi(q_t) \cdot \Phi(e_i)}{\|\Phi(q_t)\| \|\Phi(e_i)\|} \right),
\end{equation}
where $\Phi(\cdot)$ denotes a frozen pre-trained semantic encoder (e.g., a lightweight text-embedding model), and $\sigma(\cdot)$ is a scaling function to bound the score within $[0, 1]$. 

\subsubsection{Dynamic Operational Utility}
The watermark injection is expected to be confined to agents that will not degrade real-time service quality.
To prevent the routing module from selecting overloaded or slow-responding agents, the scoring module incorporates two runtime metrics: response latency $\hat{\ell}_{i,t}$ and current workload $\hat{c}_{i,t}$, both normalized to $[0, 1]$. Guided by Service-Level Agreement (SLA) specifications, the dynamic operational utility score $s_{i,t}^{\text{ope}}$ is defined as follows:
\begin{equation}
    \label{eq:operational_utility}
    s_{i,t}^{\text{ope}} = \max \left( 0, 1 - (\omega_l \hat{\ell}_{i,t} + \omega_c \hat{c}_{i,t}) \right),
\end{equation}
where $\omega_l$ and $\omega_c$ are weighting factors that balance the relative importance of latency and computational load.

For each candidate agent $a_i \in \mathcal{A}$, the synthesized reference score $\tilde{s}_{i,t}$ is defined as a convex combination of these two metrics:
\begin{equation}
    \tilde{s}_{i,t} = w_{\text{sem}} \cdot s_{i,t}^{\text{sem}} + w_{\text{ope}} \cdot s_{i,t}^{\text{ope}},
    \label{eq:score_synthesis}
\end{equation}
The trade-off hyperparameters $w_{\text{sem}}, w_{\text{ope}} \in [0, 1]$ satisfy the constraint $w_{\text{sem}} + w_{\text{ope}} = 1$, allowing the MAS owner to balance the emphasis between semantic fidelity and operational utility.

\subsection{Adaptive Watermark Embedding}
According to the results of the scoring module, DUOW identifies the available decision slack for watermark embedding at each routing step. When the gap between the top two candidates is small, it indicates that selecting the two candidates would yield comparable performance, making the routing decision less susceptible to watermark perturbations; thereby, the current watermark bit is embedded by redirecting the routing choice from the top-ranked candidate to the second-ranked one. Otherwise, the routing step is skipped without altering the native decision, and the bit is deferred for evaluation at the next routing step. 

\subsubsection{Adaptive Redundancy Interval Determination}
Formally, the scoring module computes a score $\tilde{s}_{i,t}$ for each candidate agent $a_i$, producing a score vector $\tilde{\mathbf{s}}_t = [\tilde{s}_{1,t}, \tilde{s}_{2,t}, \dots, \tilde{s}_{n,t}] \in [0,1]^n$ over the entire candidate pool. Let $a_{\text{top1},t}$ and $a_{\text{top2},t}$ denote the top-ranked and second-ranked agents, with corresponding scores $\tilde{s}_{\text{top1},t}$ and $\tilde{s}_{\text{top2},t}$. DUOW computes the score gap as:
\begin{equation}
\Delta\tilde{s}_t = \tilde{s}_{\text{top1},t} - \tilde{s}_{\text{top2},t},
\label{eq:score_gap}
\end{equation}
where $\Delta\tilde{s}_t$ measures the score separation between the top-ranked candidate and its closest alternative, with smaller values indicating greater potential decision slack.

A straightforward approach is to establish a fixed threshold: if the score gap surpasses this threshold, the routing step is regarded as a redundancy interval and chosen for watermark embedding; otherwise, it is bypassed.
However, a fixed threshold is insufficient because it implicitly assumes that the same score gap always corresponds to the same degree of decision slack. In practice, the score gap captures only the relative competitiveness between the top-ranked candidate and its closest alternative, but does not account for the score of the top-ranked candidate, which reflects how strongly the routing policy prefers the current routing decision. Consequently, routing decisions with the same score gap may still correspond to different levels of routing preference. Specifically, among such routing decisions, a higher score of the top-ranked candidate indicates a stronger routing preference, making routing redirection more likely to affect task utility, whereas a lower score of the top-ranked candidate suggests a weaker preference and therefore greater available decision slack. As a result, the same score gap may correspond to different risks of routing redirection, making a fixed threshold inadequate for accurately identifying redundancy intervals suitable for watermark embedding.

To address this limitation, we incorporate the score of the top-ranked candidate to better estimate the available decision slack, and propose the following adaptive threshold $\tau_t$:
\begin{equation}
\tau_t = \lambda\Delta\tilde{s}_t + \alpha(1-\tilde{s}_{\text{top1},t})^2,
\label{eq:adaptive_threshold}
\end{equation}
where $\lambda\in[0,1]$ and $\alpha>0$ control the contributions of the two components. Specifically, $\Delta\tilde{s}_t$ captures the relative competitiveness between the top-ranked candidate and its closest alternative, while $(1-\tilde{s}_{\text{top1},t})^2$ adjusts the threshold according to the score of the top-ranked candidate, yielding different estimates of the available decision slack for routing decisions with identical score gaps. By combining these two components, the adaptive threshold enables redundancy interval identification to account for both candidate competitiveness and routing preference.

\begin{algorithm}[!t]
\caption{DUOW-Watermark Embedding Algorithm}
\label{alg:embedding}
\begin{algorithmic}[1]
\REQUIRE Query $q_t$, candidate pool $\mathcal{A}$, key $k$, payload $\mathcal{W} = \{b_1, \dots, b_M\}$, pointer $j$, and hyperparameters $\{\lambda, \alpha, w_{\text{sem}}, w_{\text{ope}}, \omega_l, \omega_c\}$.
\ENSURE Watermarked routing destination $a_{\text{wm},t}^*$.

\vspace{0.2em}
\item[] \hspace{-1.2em}$\triangleright$ \textbf{\textit{Multi-Criteria Score Synthesis}}
\FOR{each candidate agent $a_i \in \mathcal{A}$}
    \STATE Synthesize global score $\tilde{s}_{i,t} = w_{\text{sem}} \cdot s_{i,t}^{\text{sem}} + w_{\text{ope}} \cdot s_{i,t}^{\text{ope}}$;
\ENDFOR

\vspace{0.2em}
\item[] \hspace{-1.2em}$\triangleright$ \textbf{\textit{Boundary Demarcation}}
\STATE Identify top two candidates $a_{\text{top1},t}$ and $a_{\text{top2},t}$ with the highest and second-highest scores, respectively;
\STATE $\Delta \tilde{s}_t = \tilde{s}_{\text{top1},t} - \tilde{s}_{\text{top2},t}; \quad \tau_t = \lambda \cdot \Delta \tilde{s}_t + \alpha \cdot (1 - \tilde{s}_{\text{top1},t})^2$;

\vspace{0.2em}
\item[] \hspace{-1.2em}$\triangleright$ \textbf{\textit{Trajectory Redirection \& Modulation}}
\IF{$\Delta \tilde{s}_t < \tau_t$}
    \STATE $\Omega_t \leftarrow \{a_{\text{top1},t}, a_{\text{top2},t}\}; \quad (a_t^{(0)}, a_t^{(1)}) \leftarrow \text{Sort}_{\text{UUID}}(\Omega_t)$;
    \STATE $r_t = \text{LSB}(\text{Hash}(k \parallel t)); \quad a_{\text{wm},t}^* = a_t^{(b_j \oplus r_t)}$;
    \STATE Advance and wrap watermark pointer: $j \leftarrow (j \bmod M) + 1$;
\ELSE
    \STATE Maintain native choice and defer insertion: $a_{\text{wm},t}^* = a_{\text{top1},t}$;
\ENDIF
\RETURN $a_{\text{wm},t}^*$
\end{algorithmic}
\end{algorithm}

\subsubsection{Trajectory Redirection by Watermarks}
With the redundancy intervals determined, we now embed the watermark bits into the routing decisions at those intervals. A redirection refers to the operation of switching the routing choice from the top candidate to the alternative one during this embedding process.

The payload is a binary sequence $\mathcal{W} = \{b_1, b_2, \dots, b_M\}$, embedded sequentially under the control of a dynamic pointer $j$. To ensure consistent recovery across platforms, the candidate pair is canonically ordered by UUID before redirection:
\begin{equation}
    \mathcal{S}_t^{\text{sort}} = \text{Sort}(\{a_{\text{top1},t}, a_{\text{top2},t}\}) = (a_t^{(0)}, a_t^{(1)}),
    \label{eq:canonical_sort}
\end{equation}
where $a_t^{(0)}$ and $a_t^{(1)}$ are mapped to logical slots 0 and 1, respectively. The final routing decision is then:
\begin{equation}
    a_{\text{wm},t}^* = 
    \begin{cases}
        a_{\text{top1},t}, & \text{if } \Delta \tilde{s}_t \ge \tau_t, \\
        a_t^{(b_j \oplus r_t)}, & \text{if } \Delta \tilde{s}_t < \tau_t.
    \end{cases}
    \label{eq:embedding_decision}
\end{equation}
If $\Delta \tilde{s}_t < \tau_t$, the current bit $b_j$ is embedded by redirecting the choice from the top candidate to the canonical slot determined by $b_j \oplus r_t$ and $j$ advances to the next bit. Otherwise, the decision remains unchanged and the bit is deferred to the next viable interval. The complete workflow is formalized in Algorithm~\ref{alg:embedding}.

\subsection{Watermark Extraction and Verification}
The verifier is able to perform blind watermark extraction without access to the suspect model's internal states. The verifier shares the scoring module and the private key $k$ with the embedding phase. From the suspect system's public API, the verifier obtains the decision log $\mathcal{L}$, which records the sequence of routing decisions made by the suspect model. The extraction process then reconstructs the watermark by first locating the carrier steps that satisfy the embedding boundary condition, recovering the individual bits from those steps, and finally aggregating them via statistical consensus to tolerate partial log loss.

\subsubsection{Carrier Localization and Bit Extraction}
For each step $t$ in the collected log $\mathcal{L}$, the verifier computes candidate scores using the same scoring module as in embedding, identifies the top two candidates $a_{\text{top1},t}$ and $a_{\text{top2},t}$, and computes the score gap $\Delta \tilde{s}_t$. A step is identified as a carrier if it satisfies:
\begin{equation}
    \Delta \tilde{s}_t = \tilde{s}_{\text{top1},t} - \tilde{s}_{\text{top2},t} < \tau_{t},
    \label{eq:extraction_boundary}
\end{equation}
where $\tau_t$ is recomputed from $\lambda$ and $\alpha$. This filters out deterministic steps where no watermark could have been embedded.

For each carrier step $t$, the verifier sorts the pair $\{a_{\text{top1},t}, a_{\text{top2},t}\}$ by UUID, replicating the canonical reference sequence $\mathcal{S}_t^{\text{sort}} = (a_t^{(0)}, a_t^{(1)})$ from Eq.~(\ref{eq:canonical_sort}). It then determines whether the logged decision $a_{\text{wm},t}^*$ matches $a_t^{(0)}$ or $a_t^{(1)}$, i.e., $a_{\text{wm},t}^* = a_t^{(s)}$ for some $s \in \{0,1\}$. Using the private key $k$ to reproduce the same pseudo-random mask $r_t = \text{LSB}(\text{Hash}(k \parallel t))$ used during embedding, the verifier recovers the embedded bit:
\begin{equation}
    b'_t = s \oplus r_t.
    \label{eq:bit_decoding}
\end{equation}

\subsubsection{Redundant Voting Consensus}
Since decision trace logs $\mathcal{L}$ may be partially erased or truncated in practice, preventing complete watermark recovery during verification, DUOW applies a repetition-based encoding strategy before embedding: the watermark sequence $\mathcal{W}$ is repeatedly and cyclically inserted along the routing trajectory. This comes at the cost of additional capacity, but MAS tasks typically involve long decision chains with sufficient redundant routing steps to accommodate the repeated payload.

During verification, candidate bits $b'_t$ extracted from all carrier steps are aggregated into voting pools according to their periodic position:
\begin{equation}
    \mathcal{V}_j = \{ b'_t \mid t \equiv j \pmod{M} \}.
    \label{eq:voting_pool}
\end{equation}
For each payload index $j$, the value of $\hat{b}_j$ is determined by majority vote:
\begin{equation}
    \hat{b}_j = \arg\max_{v \in \{0,1\}} \sum_{b' \in \mathcal{V}_j} \mathbf{1}(b' = v).
    \label{eq:majority_voting}
\end{equation}
The formula finds which bit value $v \in \{0,1\}$ receives the most votes among the recovered copies in $\mathcal{V}_j$. This ensures that even when part of the log is lost, the surviving copies are sufficient to recover the complete payload through statistical consensus. The complete workflow is formalized in Algorithm~\ref{alg:extraction}.

\begin{algorithm}[!t]
\caption{DUOW Ownership Verification Algorithm}
\label{alg:extraction}
\begin{algorithmic}[1]
\REQUIRE Log $\mathcal{L}$, candidate pool $\mathcal{A}$, key $k$, payload length $M$, and hyperparameters $\{\lambda, \alpha, w_{\text{sem}}, w_{\text{ope}}, \omega_l, \omega_c\}$.
\ENSURE Consensus payload $\mathcal{W} = \{b_1, b_2,\dots, b_M\}$

\vspace{0.2em}
\item[] \hspace{-1.2em}$\triangleright$ \textbf{\textit{Boundary Reconstruction \& Carrier Identification}}
\STATE Initialize tracking pools $\mathcal{V}_j \leftarrow \emptyset$ for $j = 1, \dots, M$;
\FOR{each routing step $t$ recorded in log $\mathcal{L}$}
    \STATE Re-evaluate synthesized scores $\tilde{s}_{i,t}$ for all $a_i \in \mathcal{A}$;
    \STATE Identify top two candidates $a_{\text{top1},t}$ and $a_{\text{top2},t}$ with the highest and second-highest scores, respectively;
    \STATE $\Delta \tilde{s}_t = \tilde{s}_{\text{top1},t} - \tilde{s}_{\text{top2},t}; \quad \tau_t = \lambda \cdot \Delta \tilde{s}_t + \alpha \cdot (1 - \tilde{s}_{\text{top1},t})^2$;
    \IF{$\Delta \tilde{s}_t < \tau_t$}

        \vspace{0.2em}
        \item[] \hspace{-1.2em}$\triangleright$ \textbf{\textit{Logical Slot Replication \& Bit Recovery}}
        \STATE $\Omega_t \leftarrow \{a_{\text{top1},t}, a_{\text{top2},t}\};$
        \STATE $(a_t^{(0)}, a_t^{(1)}) \leftarrow \text{Sort}_{\text{UUID}}(\Omega_t);$
        \STATE $r_t = \text{LSB}(\text{Hash}(k \parallel t))$;
        \STATE Determine $s \in \{0, 1\}$ such that $a_{\text{wm},t}^* = a_t^{(s)}$;
        \STATE Recover embedded bit: $b'_t = s \oplus r_t$;

        \vspace{0.2em}
        \item[] \hspace{-1.2em}$\triangleright$ \textbf{\textit{Redundant Voting Consensus}}
        \STATE $\mathcal{V}_{(t-1) \bmod M + 1} \leftarrow \mathcal{V}_{(t-1) \bmod M + 1} \cup \{b'_t\}$;
    \ENDIF
\ENDFOR
\FOR{each payload index $j = 1, \dots, M$}
    \STATE Decode consensus bit $\hat{b}_j$ via majority voting (Eq.~\ref{eq:majority_voting});
\ENDFOR
\RETURN $\mathcal{W} = \{b_1, b_2,\dots, b_M\}$
\end{algorithmic}
\end{algorithm}

\section{Experiments}
\subsection{Experimental Setup}
\subsubsection{Environments and Tasks}
We evaluate DUOW on three multi-agent dispatching benchmarks:
\begin{itemize}
    \item \textbf{ToolBench:} A multi-step tool-calling benchmark with sequential routing over diverse tool configurations.
    \item \textbf{DBBench:} A database routing benchmark that requires high-fidelity query execution trajectories.
    \item \textbf{Gorilla:} An API dispatching benchmark with two variants, Gorilla-HF (HuggingFace) and Gorilla-TF (TensorFlow Hub), simulating developer toolchain selection.
\end{itemize}

\subsubsection{Experimental Configuration}
The evaluation system consists of two core components:
\begin{itemize}
    \item \textbf{Routing Agent $R$:} Implemented using Gemma-3-12B-it \cite{gemmateam2025gemma3technicalreport}, which executes the native routing policy $\pi_0(q_t)$ to map each subtask query $q_t$ to a baseline routing action $a_t$.
    \item \textbf{DUOW Tool:} A lightweight plugin that uses the \texttt{bge-small-zh-v1.5} model as the text-embedding backbone. It computes dense vector representations of queries and agent descriptions to provide semantic utility scores for boundary evaluation.
\end{itemize}

\subsubsection{Baselines}
We compare DUOW against AgentMark, a state-of-the-art multi-agent watermarking method, in terms of performance trade-offs and computational overhead. The unwatermarked routing performance (Base) serves as the reference for utility evaluation.

\begin{figure*}[!t]
    \centering

    \subfloat[ToolBench]{\includegraphics[width=0.48\linewidth, trim=2 2 2 2, clip]{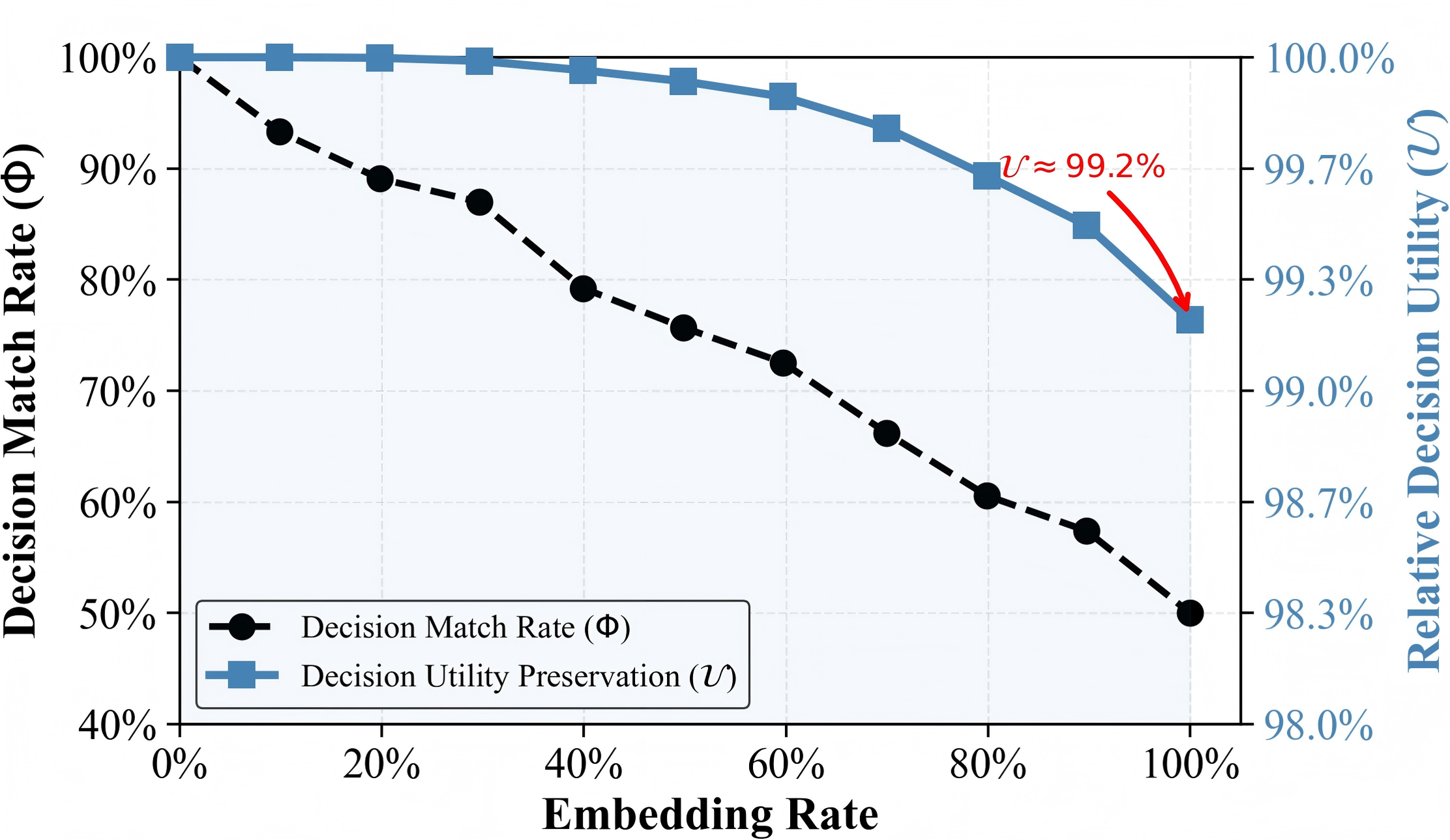}%
    \label{fig_1a}}
    \hfil
    \subfloat[DBBench]{\includegraphics[width=0.48\linewidth, trim=2 2 4 2, clip]{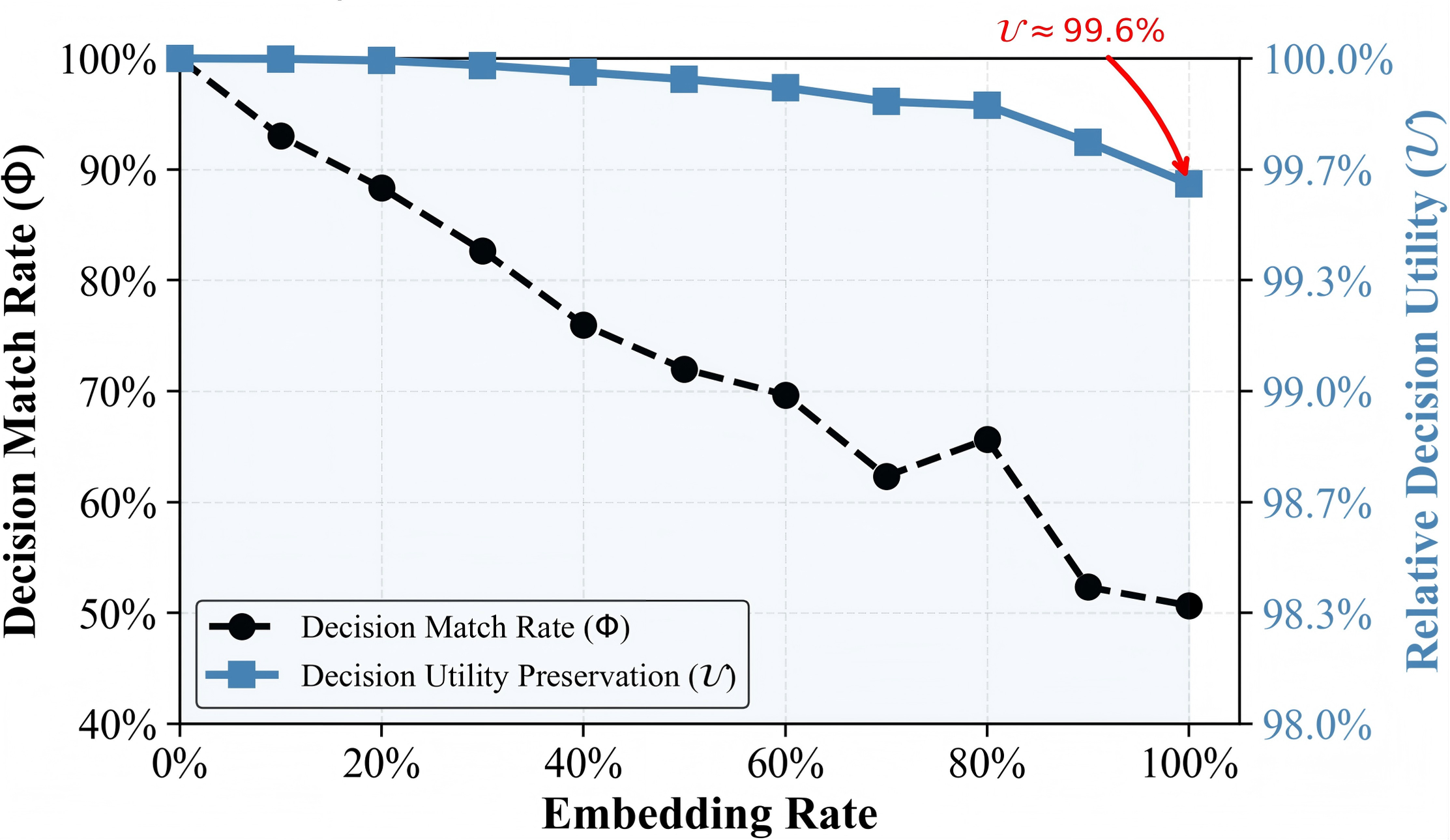}%
    \label{fig_1b}}
    
    \vspace{0.2cm}

    \subfloat[Gorilla-HF]{\includegraphics[width=0.48\linewidth, trim=2 4 2 2, clip]{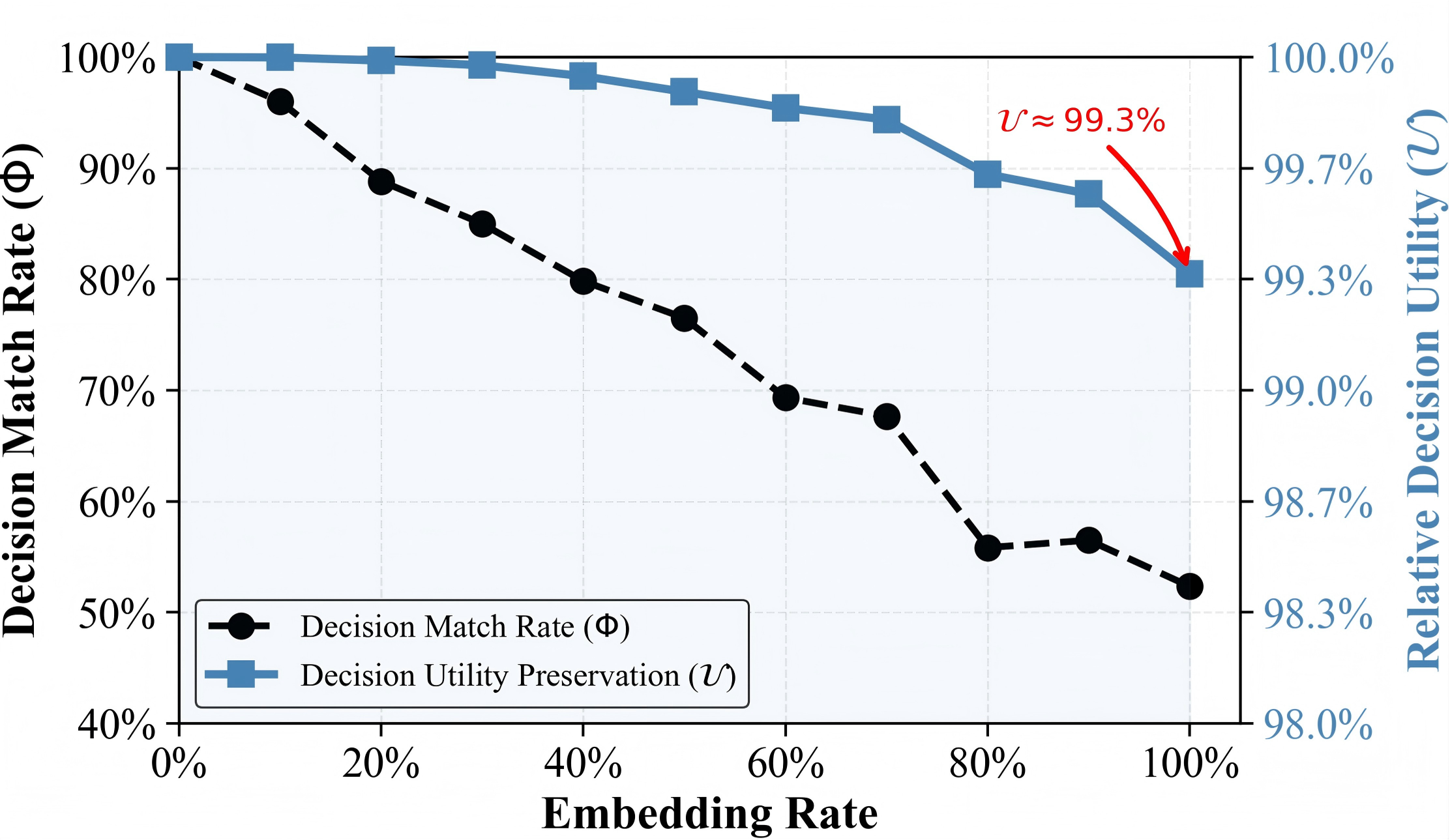}%
    \label{fig_1c}}
    \hfil
    \subfloat[Gorilla-TF]{\includegraphics[width=0.48\linewidth, trim=2 2 2 2, clip]{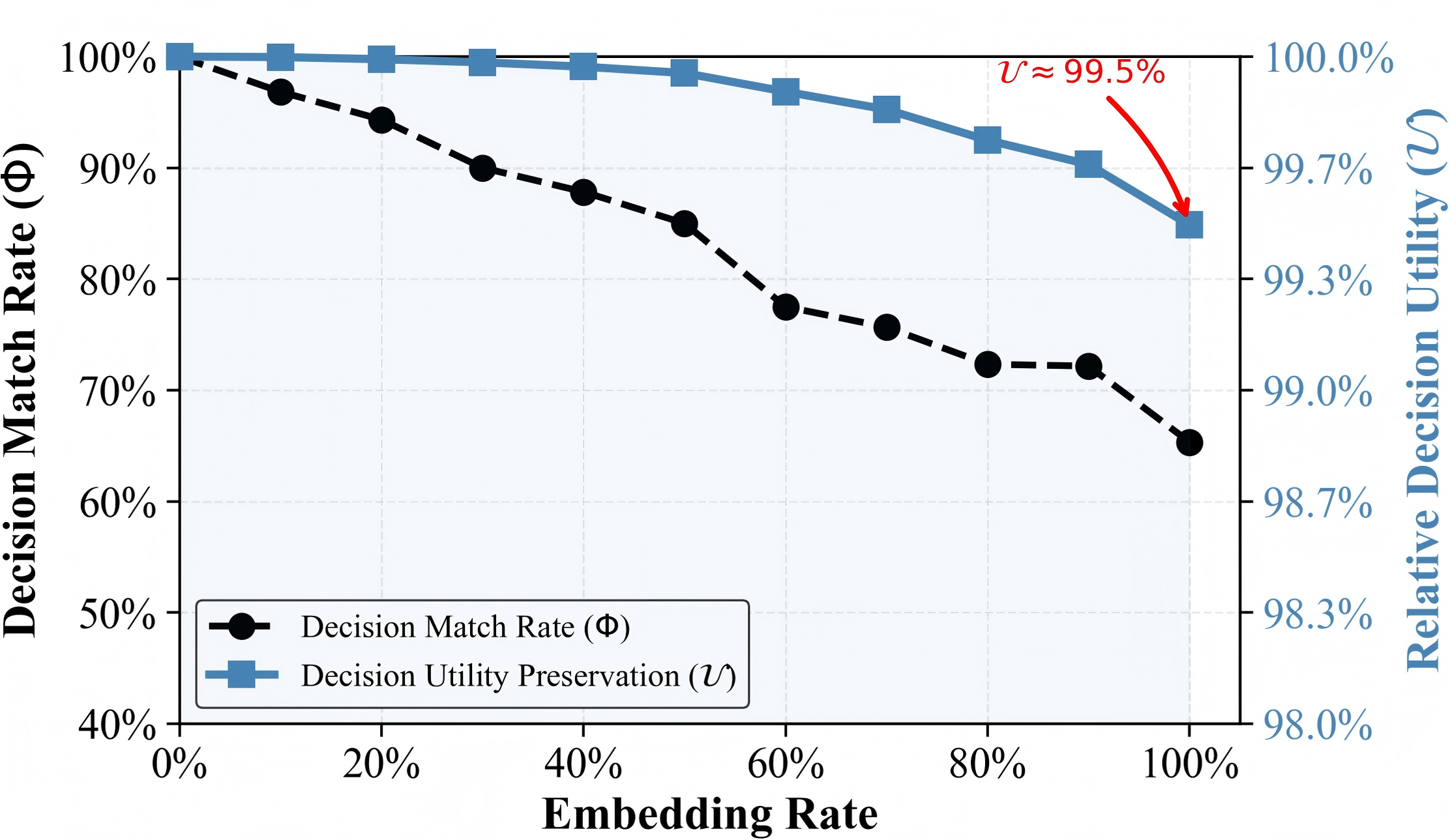}%
    \label{fig_1d}}
    
    \caption{Influence of watermark embedding rates on decision match rate $\Phi$ and relative decision utility $U$ across four heterogeneous multi-agent routing benchmarks: (a) ToolBench, (b) DBBench, (c) Gorilla-HF, and (d) Gorilla-TF. The decision match rate $\Phi$ (represented by the black dashed line with circular markers) quantifies the behavioral alignment between the watermarked routing policy and the unwatermarked baseline (Base), while the relative decision utility $U$ (represented by the blue solid line with square markers) reflects the preservation percentage of task-execution performance relative to Base. As the embedding rate approaches $100\%$, DUOW systematically maintains near-lossless macro-level utility ($U \ge 99.2\%$) across all scenarios while ensuring that the watermark activation rate ($\Phi$) scales proportionally with the target embedding density.}
    \label{fig_1}
\end{figure*}

\subsection{Feasibility Analysis of Slack-Based Redirection}
\label{subsec:feasibility}
To verify that redirecting routing decisions among functionally equivalent candidates does not significantly compromise task performance, we conduct experiments on ToolBench, DBBench, Gorilla-TF, and Gorilla-HF. For each benchmark, we embed watermarks at embedding rates ranging from 0\% to 100\%, and compare the watermarked routing trajectories against the native baseline. We evaluate two aspects: how often the watermark alters the routing choice, and whether such deviations affect the final task utility. These are quantified by the following two metrics:
\begin{itemize}
    \item \textbf{Decision Match Rate $\Phi$}: For each routing step, we compare the agent selected by the watermarked policy against the native baseline choice. $\Phi$ is the fraction of steps where the two match, so a lower $\Phi$ indicates more frequent watermark-induced redirection.
    \item \textbf{Relative Decision Utility $U$}: For each task, we verify whether it is successfully completed under the watermarked policy and compare this against the baseline success rate. $U$ is the ratio of the watermarked success rate to the baseline success rate, reflecting how much of the original task performance is preserved.
\end{itemize}

Fig.~\ref{fig_1} shows the results across all four benchmarks. As the embedding rate increases, the decision match rate $\Phi$ decreases monotonically from the baseline, converging to approximately 50\% at full saturation, for example, 50.8\% for ToolBench and 53.0\% for DBBench. This decline is expected: as more bits are embedded, a larger proportion of routing decisions are redirected from the native choice to an equivalent alternative, naturally lowering the match rate with the baseline. Notably, the convergence values are slightly above 50\%, such as 53.0\% for DBBench and 52.4\% for Gorilla-TF.

More importantly, despite the increasing divergence in routing choices, relative decision utility ($U$) remains remarkably stable across all benchmarks. Even under the maximum embedding rate, task fidelity maintains a tight lower bound of 99.2\% for ToolBench and 99.6\% for DBBench, with Gorilla-HF and Gorilla-TF achieving 99.3\% and 99.2\%, respectively. This stability reflects the effectiveness of two key components: the scoring module identifies functionally equivalent candidate pairs by jointly considering semantic alignment and runtime operational utility, while the adaptive threshold ensures that redirection is applied only when the top two candidates are sufficiently close in capability. Together, they confine watermark-induced redirections to truly equivalent alternatives and prevent performance degradation even over long task horizons. These results collectively demonstrate that DUOW effectively exploits decision slack to accommodate watermark payloads while preserving overall task utility.

\subsection{Ablation Studies}
The adaptive threshold $\tau_t$, as defined in Section~\ref{sec:Methods}, determines the condition under which watermark embedding is permitted: $\lambda$ regulates sensitivity to the score gap, while $\alpha$ penalises low confidence in the top-ranked candidate. To evaluate the impact of these two hyperparameters, we conduct a grid search over $\lambda \in [0, 1.0]$ and $\alpha \in [0, 1.0]$ with a step size of 0.1. For each configuration, we run the watermark embedding process on ToolBench, DBBench, Gorilla-TF, and Gorilla-HF and record both the achieved embedding rate and the resulting task utility.

\subsubsection{Hyperparameter Impact on Embedding Rate}
\begin{figure*}[!t]
    \centering
    \subfloat[Hyperparameter Impact of on Embedding Rate]{\includegraphics[width=0.48\linewidth]{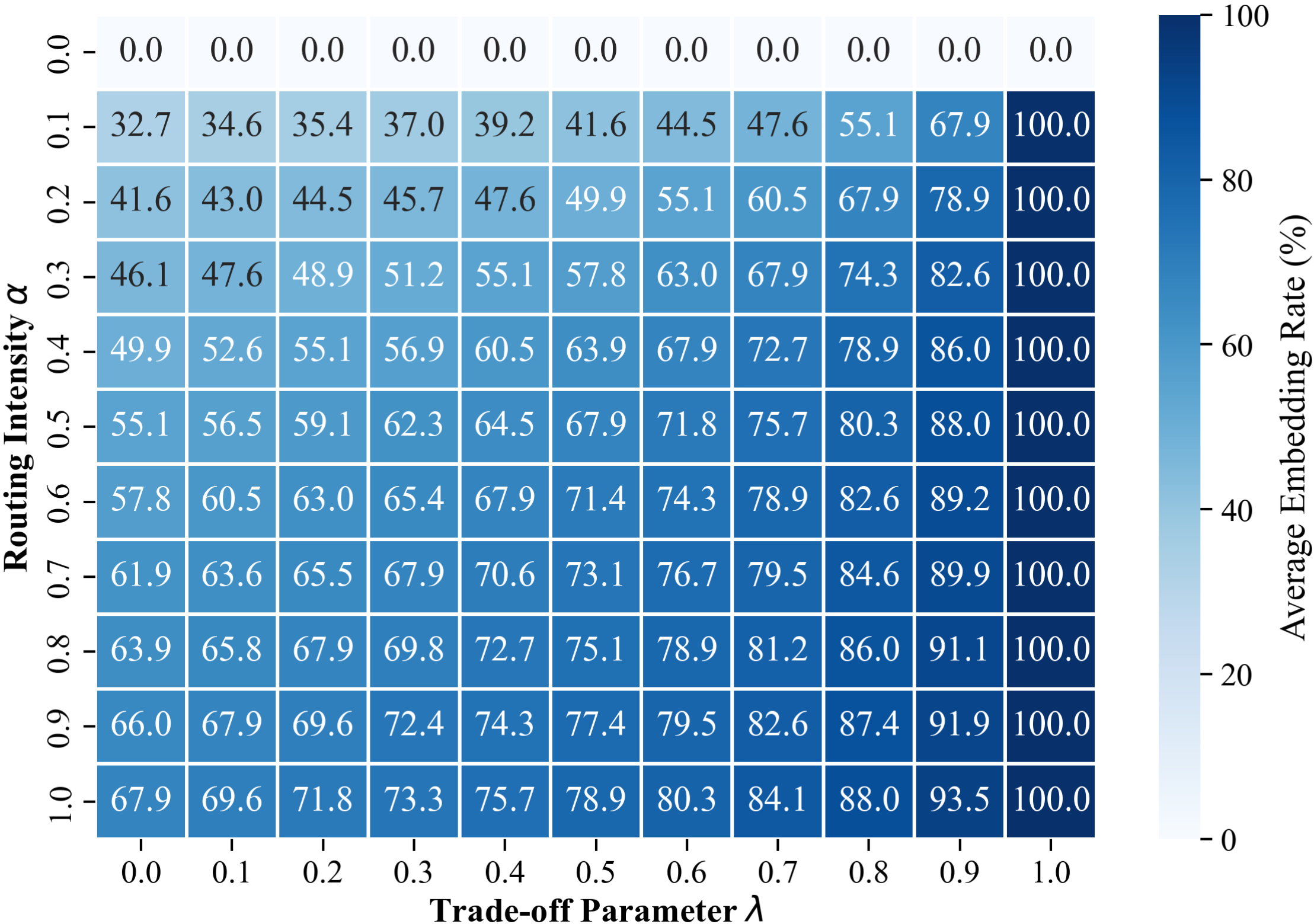}%
    \label{fig_2a}}
    \hfil
    \subfloat[Hyperparameter Impact of on Decision Utility]{\includegraphics[width=0.48\linewidth]{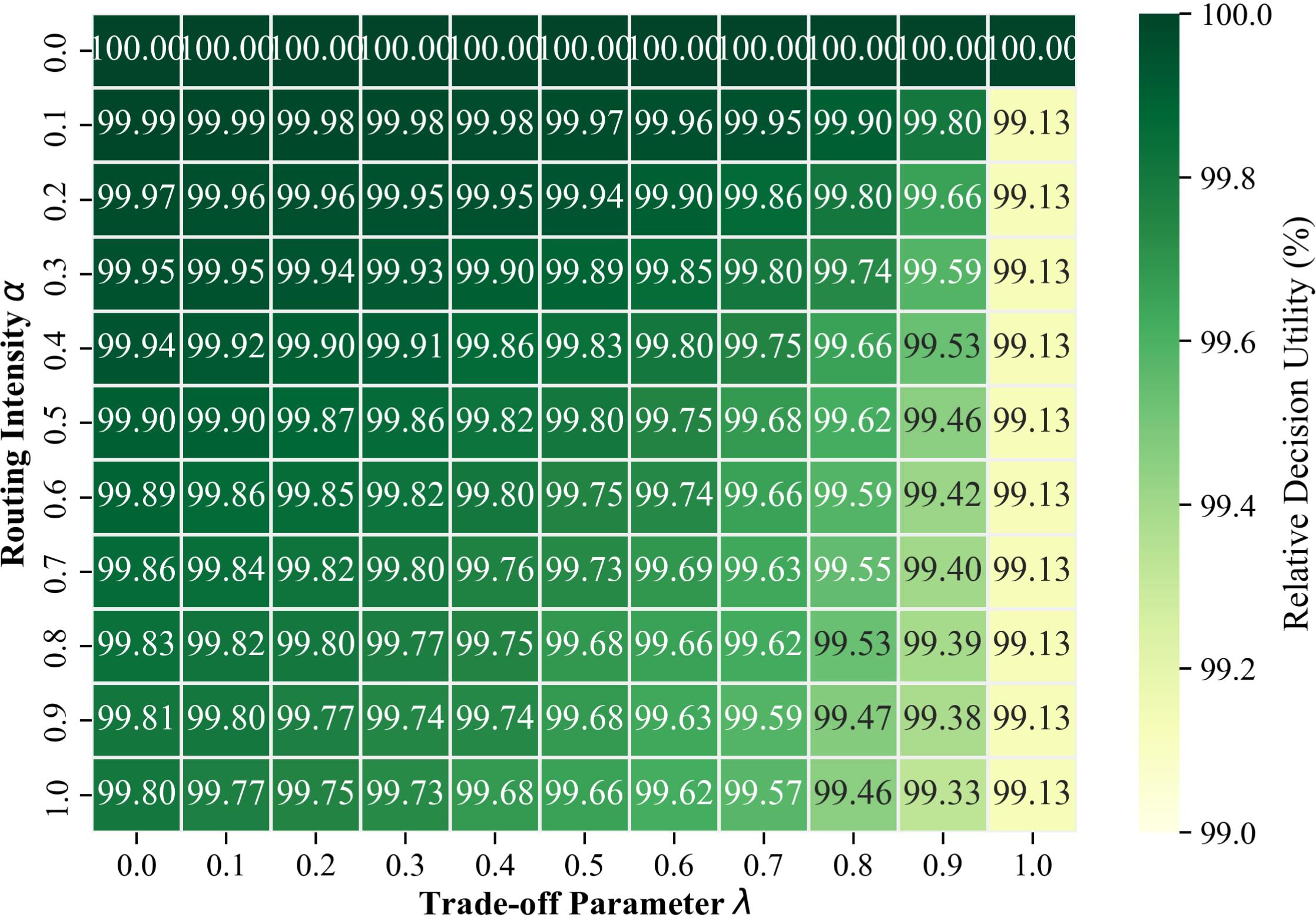}%
    \label{fig_2b}}
    \caption{Ablation of hyperparameters $\lambda$ and $\alpha$: (a) impact on average embedding rate, demonstrating continuous and predictable control over embedding capacity; and (b) impact on decision utility, showing that utility remains stable even as embedding capacity increases.}
    \label{fig_2}
\end{figure*}

To examine the effect of $\lambda$ and $\alpha$ on embedding capacity, we measure the embedding rate under different hyperparameter configurations. The results are illustrated in Fig.~\ref{fig_2a}. When both hyperparameters are set to zero, no watermark is embedded. At $\lambda = 0.5$ and $\alpha = 0.1$, the embedding rate reaches 41.6\%. Increasing to $\lambda = \alpha = 0.5$ raises the embedding rate to 67.9\% on average. At the extreme configuration $\lambda = 0.9, \alpha = 0.9$, the embedding rate reaches approximately 91.9\%. Across all four benchmarks, the embedding rate increases monotonically and smoothly with both hyperparameters. The two hyperparameters thus enable fine-grained and predictable adjustment of embedding capacity across diverse routing scenarios.

\begin{figure*}[!t]
    \centering
    \subfloat[ToolBench]{\includegraphics[width=0.48\linewidth]{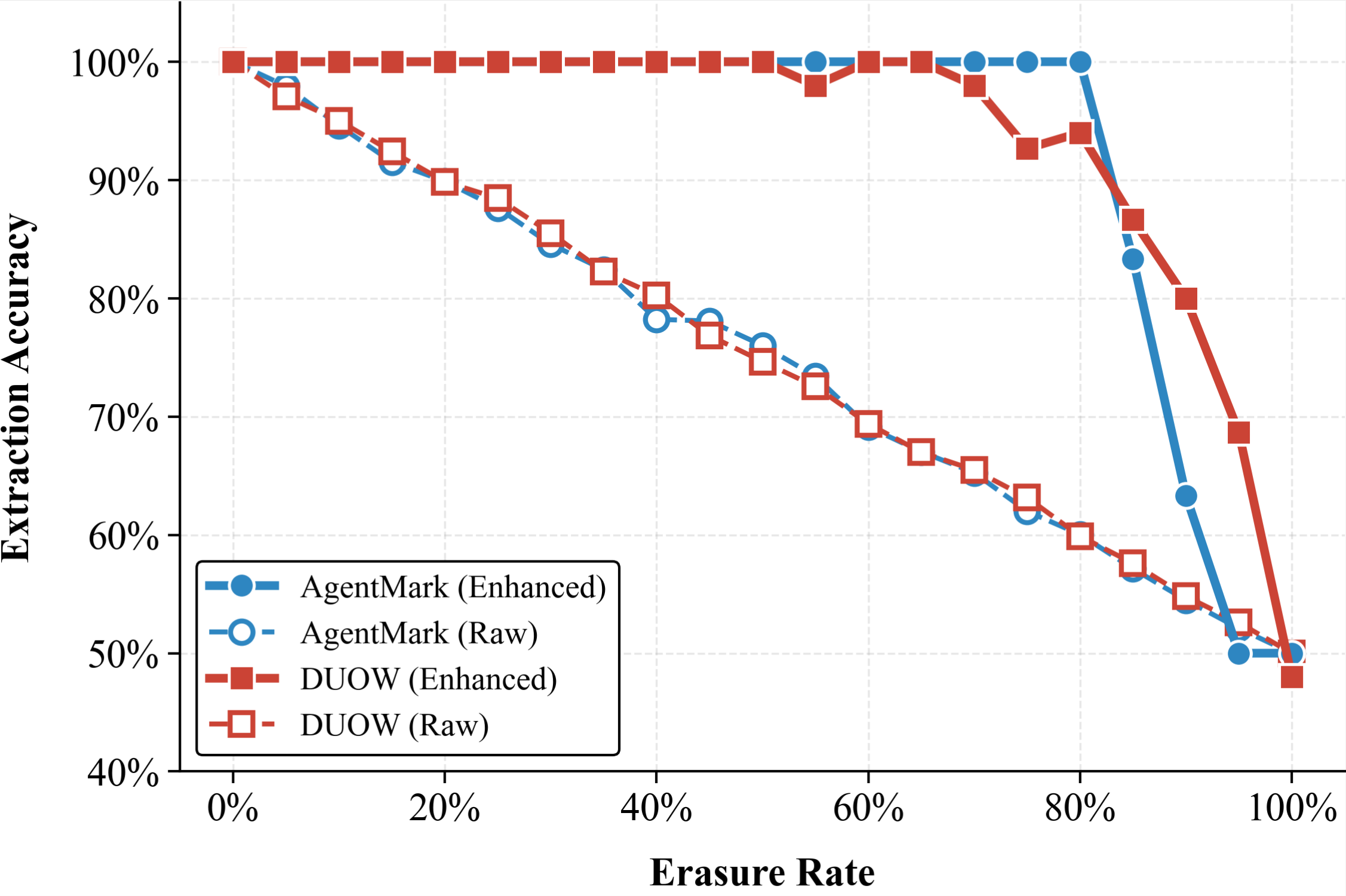}%
    \label{fig_4a}}
    \hfil
    \subfloat[DBBench]{\includegraphics[width=0.48\linewidth]{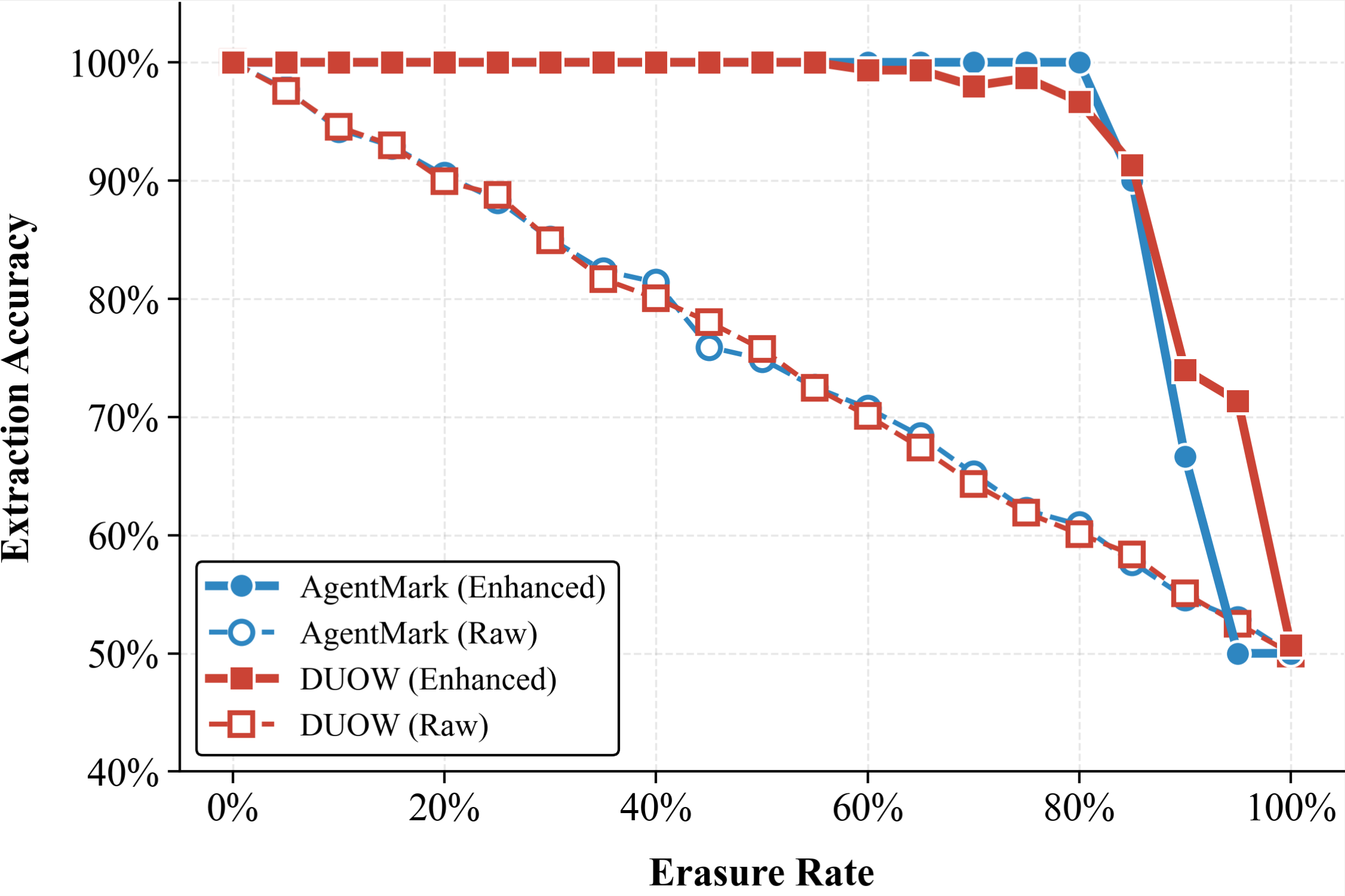}%
    \label{fig_4b}}
    \caption{Resistance to log erasure: watermark payload recovery accuracy under varying erasure rates across (a) ToolBench and (b) DBBench. The experiment evaluates extraction robustness across four operational modes: (i) AgentMark (Raw) and (ii) DUOW (Raw) represent the baseline configurations without error correction; (iii) AgentMark (Enhanced) utilizes Random Linear Network Coding (RLNC) decoding for payload recovery; and (iv) DUOW (Enhanced) deploys the statistical consensus-driven redundant voting mechanism.}
    \label{fig_4}
\end{figure*}

\subsubsection{Hyperparameter Impact on Decision Utility}
To assess whether increased embedding capacity comes at the cost of performance, we measure the relative decision utility $U$ under three representative configurations: no embedding ($\lambda = \alpha = 0$), moderate embedding ($\lambda = \alpha = 0.5$), and extreme embedding ($\lambda = \alpha = 1.0$). As shown in Fig.~\ref{fig_2b}, $U$ remains stable across all three cases. When no watermark is embedded, the utility is 100\%. At the moderate embedding rate of 67.9\%, the utility remains at 100\%. Even at the extreme configuration where the embedding rate saturates, the utility only drops to 99.13\%. The total utility loss from the lowest to the highest embedding intensity is thus bounded at 0.87\%. These results demonstrate that the adaptive threshold consistently restricts redirections to functionally similar candidates, thereby preventing performance degradation even under aggressive hyperparameter settings.

The above results show that the embedding rate can be flexibly controlled from zero to over 90\% by adjusting $\lambda$ and $\alpha$, while the decision utility remains above 99\% across all configurations with a maximum loss of only 0.87\%. This indicates that performance remains stable across the tested parameter range. To balance sufficient capacity with the preservation of near-perfect utility, we choose $\lambda = \alpha = 0.5$ as the default configuration. This setting yields an embedding rate of 67.9\% while maintaining 99.80\% utility, which is sufficient for reliable watermark delivery in our experiments.

\begin{table*}[!t]
\caption{Policy cloning resistance and verification generalization across evaluation benchmarks.}
\label{tab:evaluation_results}
\centering

\setlength{\tabcolsep}{1.4pt} 

\scriptsize 

\begin{tabular}{lllccccccccccccccccccccccccc}
\toprule
Model & Method & Metric & \multicolumn{6}{c}{Toolbench} & \multicolumn{6}{c}{DBBench} & \multicolumn{6}{c}{Gorilla-HF} & \multicolumn{6}{c}{Gorilla-TF} \\
\cmidrule(r){4-9} \cmidrule(lr){10-15} \cmidrule(lr){16-21} \cmidrule(l){22-27}
& & & 10\% & 20\% & 30\% & 40\% & 50\% & 60\% & 10\% & 20\% & 30\% & 40\% & 50\% & 60\% & 10\% & 20\% & 30\% & 40\% & 50\% & 60\% & 10\% & 20\% & 30\% & 40\% & 50\% & 60\% \\
\midrule
\multirow{9}{*}{\makecell{Qwen\\2.5-7B}} 
& \multirow{3}{*}{\makecell[l]{Agent\\Mark}}
& S-ER & \textbf{99.13} & 98.57 & \textbf{98.57} & 98.14 & \textbf{98.43} & \textbf{98.71} & 100.0 & 100.0 & 100.0 & 100.0 & 99.50 & 100.0 & 98.86 & 97.71 & 96.00 & 96.43 & 95.14 & 95.14 & 96.98 & 93.79 & 91.61 & 78.36 & 88.76 & \textbf{86.24} \\
& & U-ER & 5.63 & 5.63 & 8.45 & 8.45 & 8.45 & 8.45 & 0.00 & 0.00 & 2.30 & 1.15 & 2.30 & 1.15 & 0.00 & \textbf{1.18} & 0.00 & 1.18 & 4.71 & 5.88 & 0.00 & 4.30 & 1.08 & 2.15 & 3.23 & \textbf{6.45} \\
& & O-VR & 90.40 & 90.01 & 90.27 & 89.88 & 90.14 & 90.40 & 69.69 & 69.69 & 70.38 & 70.03 & 70.03 & 70.03 & 88.15 & 87.26 & 85.61 & 86.11 & 85.35 & 85.48 & 83.89 & 81.71 & 79.39 & 68.07 & 77.21 & \textbf{75.47} \\
\cmidrule{2-27}
& \multirow{3}{*}{DUOW} 
& S-ER & 98.86 & 98.57 & 98.29 & \textbf{98.43} & 98.14 & 98.29 & 100.0 & 100.0 & 100.0 & 100.0 & \textbf{100.0} & 100.0 & \textbf{99.43} & \textbf{98.43} & \textbf{98.00} & \textbf{97.29} & \textbf{97.29} & \textbf{96.29} & \textbf{98.66} & \textbf{96.81} & \textbf{94.46} & \textbf{92.79} & \textbf{90.94} & 65.94 \\
& & U-ER & \textbf{12.07} & \textbf{13.79} & \textbf{13.79} & \textbf{22.41} & \textbf{20.69} & \textbf{24.14} & \textbf{1.69} & \textbf{1.69} & \textbf{11.86} & \textbf{5.08} & \textbf{10.17} & \textbf{44.07} & \textbf{1.72} & 0.00 & \textbf{1.72} & \textbf{8.62} & \textbf{18.97} & \textbf{27.59} & \textbf{3.64} & \textbf{14.55} & \textbf{18.18} & \textbf{21.82} & \textbf{30.91} & 5.45 \\
& & O-VR & \textbf{92.22} & \textbf{92.08} & \textbf{91.82} & \textbf{92.61} & \textbf{92.22} & \textbf{92.61} & \textbf{77.61} & \textbf{77.61} & \textbf{79.92} & \textbf{78.38} & \textbf{79.54} & \textbf{87.26} & \textbf{91.95} & \textbf{90.90} & \textbf{90.63} & \textbf{90.50} & \textbf{91.29} & \textbf{91.03} & \textbf{90.63} & \textbf{90.86} & \textbf{88.02} & \textbf{86.79} & \textbf{85.87} & 60.83 \\
\cmidrule{2-27}
& \multirow{3}{*}{Comp.}
& $\Delta$S-ER & -0.27 & 0.00 & -0.28 & 0.29 & -0.29 & -0.42 & 0.00 & 0.00 & 0.00 & 0.00 & 0.50 & 0.00 & 0.57 & 0.72 & 2.00 & 0.86 & 2.15 & 1.15 & 1.68 & 3.02 & 2.85 & 14.43 & 2.18 & -20.30 \\
& & $\Delta$U-ER & 6.44 & 8.16 & 5.34 & 13.96 & 12.24 & 15.69 & 1.69 & 1.69 & 9.56 & 3.93 & 7.87 & 42.92 & 1.72 & -1.18 & 1.72 & 7.44 & 14.26 & 21.71 & 3.64 & 10.25 & 17.10 & 19.67 & 27.68 & -1.00 \\
& & $\Delta$O-VR & 1.82 & 2.07 & 1.55 & 2.73 & 2.08 & 2.21 & 7.92 & 7.92 & 9.54 & 8.35 & 9.51 & 17.23 & 3.80 & 3.64 & 5.02 & 4.39 & 5.94 & 5.55 & 6.74 & 8.15 & 8.63 & 18.72 & 8.66 & -14.64 \\
\midrule
\multirow{9}{*}{\makecell{Llama\\3.1-8B}} 
& \multirow{3}{*}{\makecell[l]{Agent\\Mark}}
& S-ER & \textbf{99.43} & 98.86 & \textbf{99.13} & 98.43 & \textbf{98.71} & 98.43 & 100.0 & 100.0 & 100.0 & 99.50 & 96.00 & 93.50 & \textbf{99.13} & 97.57 & 95.57 & \textbf{96.14} & 95.00 & 94.86 & \textbf{97.15} & 89.93 & 90.77 & 61.74 & 81.38 & \textbf{84.40} \\
& & U-ER & 7.04 & 8.45 & 7.04 & 12.68 & 7.04 & 9.86 & 0.00 & 0.00 & 2.30 & 2.30 & 2.30 & 6.90 & 0.00 & 0.00 & 0.00 & 2.35 & 3.53 & 8.24 & 0.00 & 0.00 & 1.08 & 1.08 & 0.00 & 6.45 \\
& & O-VR & 90.92 & 90.53 & 90.53 & 90.53 & 90.27 & 90.27 & 69.69 & 69.69 & 70.38 & 70.03 & 67.60 & 67.25 & 88.28 & 87.01 & 85.22 & \textbf{85.99} & 85.10 & 85.48 & 84.03 & 77.79 & 78.66 & 53.56 & 70.39 & \textbf{73.88} \\
\cmidrule{2-27}
& \multirow{3}{*}{DUOW} 
& S-ER & 99.29 & 98.86 & 98.71 & \textbf{99.13} & 98.57 & \textbf{98.43} & 100.0 & 100.0 & 100.0 & 100.0 & \textbf{100.0} & 100.0 & 99.43 & \textbf{98.43} & \textbf{97.86} & 94.14 & \textbf{97.14} & \textbf{96.14} & 94.30 & \textbf{95.30} & \textbf{91.44} & \textbf{91.78} & \textbf{85.07} & 66.11 \\
& & U-ER & \textbf{12.07} & \textbf{15.52} & \textbf{17.24} & \textbf{24.14} & \textbf{25.86} & \textbf{27.59} & 0.00 & \textbf{8.47} & \textbf{10.17} & \textbf{5.08} & \textbf{15.25} & \textbf{37.29} & \textbf{1.72} & \textbf{5.17} & \textbf{1.72} & \textbf{13.79} & \textbf{17.24} & \textbf{29.31} & \textbf{1.82} & \textbf{10.91} & \textbf{12.73} & \textbf{18.18} & \textbf{29.09} & \textbf{10.91} \\
& & O-VR  & \textbf{92.61} & \textbf{92.48} & \textbf{92.48} & \textbf{93.27} & \textbf{93.01} & \textbf{93.01} & \textbf{77.22} & \textbf{79.15} & \textbf{79.54} & \textbf{78.38} & \textbf{80.69} & \textbf{85.71} & \textbf{91.95} & \textbf{91.29} & \textbf{90.50} & 87.99 & \textbf{91.03} & \textbf{91.03} & \textbf{86.48} & \textbf{88.17} & \textbf{84.79} & \textbf{85.56} & \textbf{80.34} & 61.44 \\
\cmidrule{2-27}
& \multirow{3}{*}{Comp.}
& $\Delta$S-ER & -0.14 & 0.00 & -0.42 & 0.70 & -0.14 & 0.00 & 0.00 & 0.00 & 0.00 & 0.50 & 4.00 & 6.50 & 0.30 & 0.86 & 2.29 & -2.00 & 2.14 & 1.28 & -2.85 & 5.37 & 0.67 & 30.04 & 3.69 & -18.29 \\
& & $\Delta$U-ER & 5.03 & 7.07 & 10.20 & 11.46 & 18.82 & 17.73 & 0.00 & 8.47 & 7.87 & 2.78 & 12.95 & 30.39 & 1.72 & 5.17 & 1.72 & 11.44 & 13.71 & 21.07 & 1.82 & 10.91 & 11.65 & 17.10 & 29.09 & 4.46 \\
& & $\Delta$O-VR & 1.69 & 1.95 & 1.95 & 2.74 & 2.74 & 2.74 & 7.53 & 9.46 & 9.16 & 8.35 & 13.09 & 18.46 & 3.67 & 4.28 & 5.28 & 2.00 & 5.93 & 5.55 & 2.45 & 10.38 & 6.13 & 32.00 & 9.95 & -12.44 \\
\bottomrule
\end{tabular}
\end{table*}

\subsection{Robustness Analysis}
In practical MAS deployments, decision logs are routinely exposed for auditing and billing purposes, making them accessible to adversaries. Such exposure introduces two types of threats. First, an adversary may erase or truncate parts of the logs to disrupt watermark extraction. Second, an adversary may collect the exposed logs and use them to train a cloned model that mimics the victim's routing policy. Given these threats, an effective watermarking scheme must remain extractable from incomplete logs and survive distillation into cloned models. To evaluate DUOW's resilience against these two threats, we conduct two sets of experiments: log erasure resistance and policy cloning resistance.

\subsubsection{Resistance to Log Erasure}
We evaluate the robustness of watermark recovery against log erasure by randomly removing routing steps from the logged trajectories at varying erasure rates ranging from 10\% to 100\%. For each erasure rate, we measure the accuracy of payload recovery and compare DUOW against AgentMark under two configurations: raw extraction without error correction, and enhanced extraction with error correction enabled (DUOW's redundant voting consensus vs. AgentMark's RLNC).

During the raw extraction phase, DUOW and AgentMark exhibit nearly identical performance across varying erasure rates (Fig.~\ref{fig_4}). Even under 20\% erasure, both maintain approximately 90\% accuracy, and their accuracy decays steadily as the erasure rate increases. This is expected: without error correction, the surviving log fragments directly determine how many watermark bits can be recovered—erased fragments simply mean lost bits. The close alignment between the two curves also provides a fair baseline for comparing their respective error-correction mechanisms.

With error correction enabled, both frameworks show significant improvements but diverge under high erasure. AgentMark achieves full recovery when the erasure rate is below 70\%, benefiting from the algebraic decoding capacity of RLNC. Beyond this threshold, its accuracy declines steadily, dropping to around 60\% at 80\% erasure and approximately 50\% at 90\%, as the number of surviving linearly independent equations falls below the solvable limit. In contrast, DUOW's statistical consensus mechanism exhibits a more gradual degradation: it sustains approximately 90\% accuracy on ToolBench and 95\% on DBBench at 80\% erasure, and still achieves 74.65\% recovery on DBBench even under 90\% log loss. These results demonstrate that DUOW offers superior robustness in severely truncated verification environments.

\begin{table*}[!t]
\caption{Embedding latency of DUOW and AgentMark across varying embedding rates, with DUOW's overhead reduction over AgentMark.}
\label{tab:embedding_time_vertical}
\centering
\resizebox{0.85\textwidth}{!}{
\begin{tabular}{ll cccccccccc}
\toprule
\multirow{2}{*}{Benchmark} & \multirow{2}{*}{Method} & \multicolumn{10}{c}{Embedding Rate} \\ 
\cmidrule(l){3-12}
& & 10\% & 20\% & 30\% & 40\% & 50\% & 60\% & 70\% & 80\% & 90\% & 100\% \\
\midrule
\multirow{3}{*}{Toolbench} 
& AgentMark (ms) & 1.09 & 2.05 & 3.29 & 4.27 & 5.40 & 6.83 & 22.45 & 25.72 & 28.40 & 33.07 \\
& DUOW (ms) & 0.07 & 0.11 & 0.27 & 0.17 & 0.23 & 0.28 & 0.29 & 0.54 & 0.39 & 0.47 \\
& Overhead Reduction (\%) & 93.58 & 94.63 & 91.79 & 96.02 & 95.74 & 95.90 & 98.71 & 97.90 & 98.63 & 98.58 \\
\midrule
\multirow{3}{*}{DBBench} 
& AgentMark (ms) & 1.05 & 2.09 & 3.11 & 4.02 & 5.45 & 6.17 & 22.66 & 25.76 & 29.46 & 33.90 \\
& DUOW (ms) & 0.062 & 0.102 & 0.14 & 0.16 & 0.26 & 0.29 & 0.38 & 0.36 & 0.38 & 0.39 \\
& Overhead Reduction (\%) & 94.10 & 95.12 & 95.50 & 96.02 & 95.23 & 95.30 & 98.32 & 98.60 & 98.71 & 98.85 \\
\midrule
\multirow{3}{*}{Gorilla-HF} 
& AgentMark (ms) & 0.98 & 1.96 & 3.07 & 4.29 & 4.78 & 5.94 & 21.19 & 23.79 & 27.21 & 30.55 \\
& DUOW (ms) & 0.04 & 0.09 & 0.13 & 0.15 & 0.22 & 0.23 & 0.29 & 0.31 & 0.34 & 0.33 \\
& Overhead Reduction (\%) & 95.92 & 95.41 & 95.77 & 96.50 & 95.40 & 96.13 & 98.63 & 98.70 & 98.75 & 98.92 \\
\midrule
\multirow{3}{*}{Gorilla-TF} 
& AgentMark (ms) & 1.09 & 2.09 & 3.31 & 4.15 & 5.04 & 6.47 & 22.23 & 25.86 & 29.06 & 30.87 \\
& DUOW (ms) & 0.06 & 0.11 & 0.15 & 0.19 & 0.21 & 0.26 & 0.29 & 0.33 & 0.39 & 0.37 \\
& Overhead Reduction (\%) & 94.50 & 94.74 & 95.47 & 95.42 & 95.83 & 95.98 & 98.70 & 98.72 & 98.66 & 98.80 \\
\bottomrule
\end{tabular}
}
\end{table*}

\subsubsection{Resistance to Policy Cloning}
We first collect routing trajectories from both DUOW-watermarked and AgentMark-watermarked agents across varying embedding rates. These trajectories are then used as the training datasets to fine-tune \texttt{Qwen2.5-7B-Instruct} \cite{hui2024qwen2} and \texttt{Llama3.1-8B-Instruct} \cite{grattafiori2024llama} as student models, aiming to replicate the watermarking behavior of the respective teacher models. Three metrics are used to quantify cloning resistance: Seen Extraction Rate (S-ER) on training-distribution tasks, Unseen Extraction Rate (U-ER) on unseen tasks, and Overall Verification Rate (O-VR) as the aggregate measure.

As reported in Table~\ref{tab:evaluation_results}, both schemes achieve S-ER above 95\% across all models, confirming that the extracted watermark behaviors can be reproduced by students on seen tasks. The key difference lies in U-ER. AgentMark shows consistently low extraction on unseen samples across most configurations. For instance, with \texttt{Llama3.1} on ToolBench at 60\% embedding, AgentMark achieves only 9.86\% U-ER, and on DBBench under the same setting, it drops to 6.90\%, indicating that its watermark signals exhibit limited transferability beyond the observed training trajectories. DUOW, in contrast, maintains substantially higher U-ER performance. Under the same 60\% embedding configuration with \texttt{Llama3.1}, DUOW reaches 27.59\% on ToolBench and 37.29\% on DBBench, and its U-ER consistently increases with the embedding rate. This cross-domain resilience translates into higher O-VR across most settings. DUOW achieves 92.61\% O-VR on ToolBench with \texttt{Llama3.1}, compared to AgentMark's 90.92\%, and delivers a maximum improvement of 32.00\% ($\Delta$O-VR) on Gorilla-TF.

This resistance advantage stems from the fundamental difference in watermark design. AgentMark relies on step-level algebraic decoding, which requires precise synchronization with local trajectory patterns. Consequently, trajectory variations introduced during policy imitation may disrupt the alignment between embedded signals and verification procedures, leading to reduced extraction performance on unseen tasks. In contrast, DUOW embeds ownership signals into routing preferences by redirecting decisions only among functionally equivalent agents. Since these redirected decisions remain consistent with normal task execution and preserve utility, they are less distinguishable from legitimate routing behaviors during imitation. As a result, student models trained on DUOW-watermarked trajectories exhibit higher transferability of watermark behaviors across unseen tasks, leading to improved verification performance compared with existing approaches.

\begin{figure}[t]
    \centering
    \includegraphics[width=\linewidth]{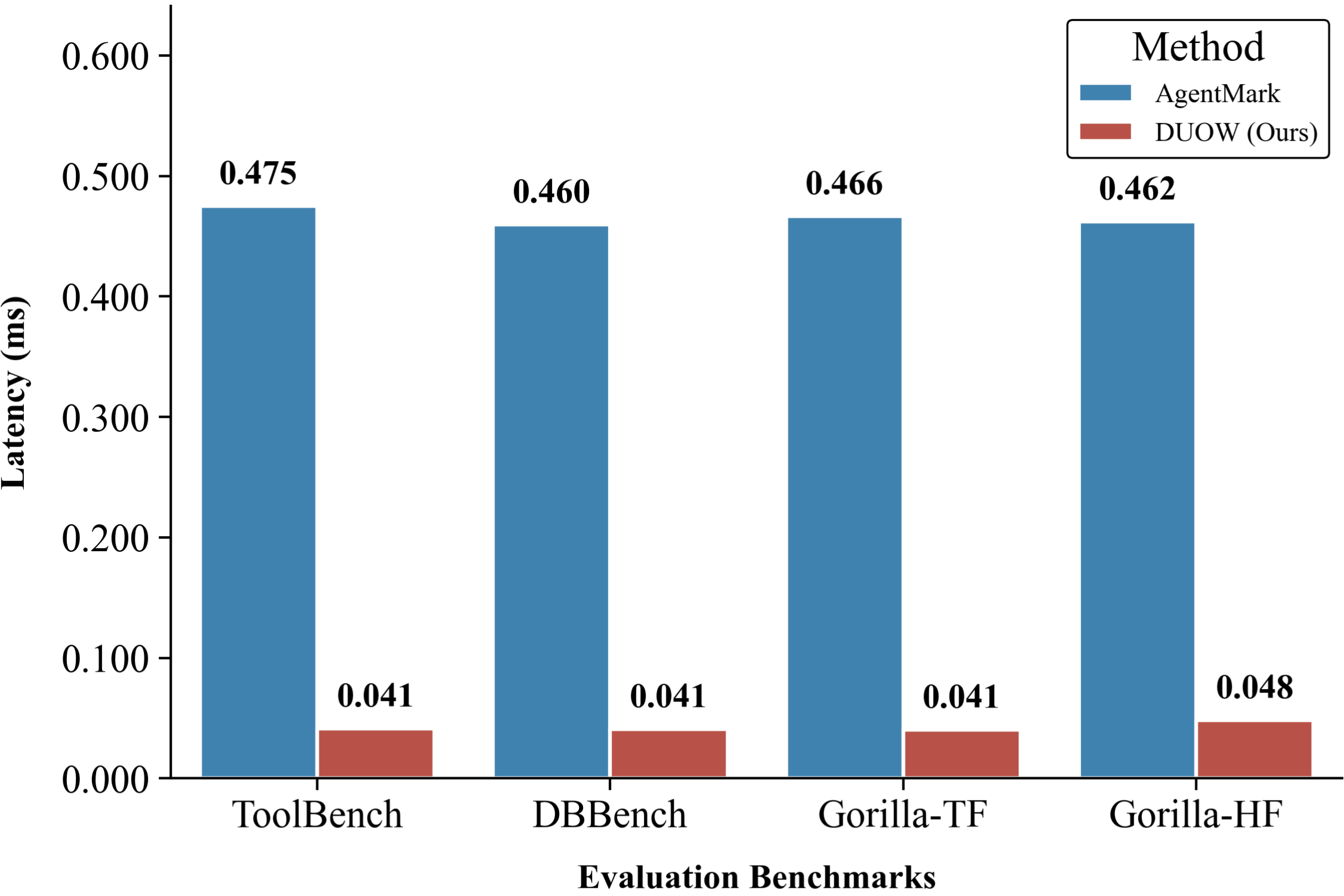}
    \caption{Computational efficiency: per-step latency comparison between DUOW and AgentMark across ToolBench, DBBench, Gorilla-TF, and Gorilla-HF.}
    \label{fig_5}
\end{figure}

\subsection{Computational Efficiency and Complexity Analysis}
DUOW's embedding latency is measured on ToolBench, DBBench, Gorilla-HF, and Gorilla-TF, and compared against AgentMark under the same settings. For each benchmark and each embedding rate, we run multiple queries and report the average latency at two granularities: the per-decision overhead and the cumulative time under different embedding rates.

Fig.~\ref{fig_5} compares the per-step latency. DUOW averages 0.043~ms per routing decision, which is less than one-tenth of AgentMark's cost of approximately 0.46~ms to 0.48~ms. The efficiency gap is structural: AgentMark requires intensive perturbation computations during each sampling step, creating a processing bottleneck. DUOW, in contrast, only computes a few scalar scores and performs simple comparisons per routing step, without any dependence on the underlying LLM. This keeps the overhead consistently low and independent of the model size.
Table~\ref{tab:embedding_time_vertical} further reports the cumulative embedding time across different embedding rates. AgentMark latency grows from 1.09~ms at 10\% embedding to 33.07~ms at 100\% saturation on ToolBench, with a notable jump from 6.83~ms at 60\% to 22.45~ms at 70\%. DUOW maintains consistently low overhead across all configurations, with total embedding time ranging from 0.04~ms to 0.54~ms across all benchmarks and embedding rates, showing no correlation with payload scaling. This translates to a peak reduction of 98.92\% on Gorilla-HF.

\section{Conclusion}
This paper presented DUOW, a decision-layer watermarking framework that secures multi-agent routing policies by encoding ownership into functionally redundant candidate selections. Comprehensive evaluations across four benchmarks confirm that DUOW preserves task utility ($\leq$99.2\%), sustains recovery under 90\% log erasure, and surpasses existing methods in cloning resistance, while incurring negligible per-step latency. Despite these merits, the framework carries inherent limitations that motivate future inquiry.
\textbf{i)Work limitations:} DUOW’s embedding capacity is inherently tied to the degree of functional overlap among candidate agents for each query and context. This overlap is not guaranteed but varies unpredictably: some steps offer multiple comparable choices, others have a single unique optimum, and worst‑case trajectories may present no embeddable opportunities at all.  
In domains with sharply defined functional boundaries, redundancy becomes scarce, rendering DUOW ineffective and copyright verification unreliable.
\textbf{ii) Future work:} To move beyond incidental redundancy, future work should pursue active redundancy engineering—for instance, via multi‑objective reinforcement learning that explicitly constructs multiple optimal solutions while preserving utility. Alternatively, cross‑step coding schemes could distribute watermark bits over entire trajectory statistics, enabling extraction even when per‑step redundancy is sparse. Ultimately, watermark protection should be internalized as an intrinsic property of the routing policy, rather than relying on environmental chance.


\bibliographystyle{IEEEtran}
\bibliography{reference.bib}

\end{document}